\documentclass[12pt,preprint]{aastex}

\usepackage{amsmath}
\usepackage{graphicx}
\usepackage{lscape}
\usepackage{color}
\shorttitle{WATER IN PROTOPLANETARY DISKS}
\shortauthors{Furuya et al.}

\begin{document}

\title{WATER IN PROTOPLANETARY DISKS: DEUTERATION AND TURBULENT MIXING}


\author{Kenji Furuya\altaffilmark{1}, Yuri Aikawa\altaffilmark{1}, Hideko Nomura\altaffilmark{2, 3, 4}, Franck Hersant\altaffilmark{5, 6}, Valentine Wakelam\altaffilmark{5, 6}}
\email{furuya@stu.kobe-u.ac.jp}

\altaffiltext{1}{Department of Earth and Planetary Sciences, Kobe University, Kobe 657-8501, Japan}
\altaffiltext{2}{Department of Astronomy, Graduate School of Science, Kyoto University, Kyoto 606-8502, Japan}
\altaffiltext{3}{National Astronomical Observatory of Japan, Osawa, Mitaka, Tokyo 181-8588, Japan}
\altaffiltext{4}{Department of Earth and Planetary Sciences, Tokyo Institute of Technology, Tokyo 152-8551, Japan}
\altaffiltext{5}{Univ. Bordeaux, LAB, UMR 5804, F-33270, Floirac, France}
\altaffiltext{6}{CNRS, LAB, UMR 5804, F-33270, Floirac, France}

%\altaffiltext{3}{CNRS and Universit\'e de Bordeaux, Observatoire Aquitain des Sciences de l'Univers, 2 rue de l'Observatoire, B.P. 89, F-33271 Floirac, France}
%\altaffiltext{4}{CNRS, LAB, UMR 5804, F-33270, Floirac, France}

%\altaffiltext{1}{Visiting Astronomer, Cerro Tololo Inter-American Observatory.
%CTIO is operated by AURA, Inc.\ under contract to the National Science
%\altaffiltext{2}{Society of Fellows, Harvard University.}
%\altaffiltext{3}{present address: Center for Astrophysics,
%    60 Garden Street, Cambridge, MA 02138}
%\altaffiltext{4}{Visiting Programmer, Space Telescope Science Institute}
%\altaffiltext{5}{Patron, Alonso's Bar and Grill}


\begin{abstract}
We investigate water and deuterated water chemistry in turbulent protoplanetary disks.
Chemical rate equations are solved with the diffusion term, mimicking turbulent mixing in vertical direction.
%Our detailed gas-grain network includes photoreactions both in the gas phase and on grain surfaces by stellar, X-ray induced, and cosmic-ray induced ultra violet photons.
Water near the midplane is transported to the disk atmosphere by turbulence and destroyed by photoreactions to produce atomic oxygen, while the atomic oxygen is transported to 
the midplane and reforms water and/or other molecules.
We find that this cycle significantly decreases column densities of water ice at $r\lesssim30$ AU, where dust temperatures are too high to reform water ice effectively.
The radial extent of such region depends on the desorption energy of atomic hydrogen.
Our model indicates that water ice could be deficient even outside the sublimation radius.
Outside this radius, the cycle decreases the D/H ratio of water ice from $\sim \!\! 2\times10^{-2}$, which is set by the collapsing core model, 
to $10^{-4}$--$10^{-2}$ in 10$^6$ yr, without significantly decreasing the water ice column density.
The resultant D/H ratios depend on the strength of mixing and the radial distance from the central star.
Our finding suggests that the D/H ratio of cometary water ($\sim$10$^{-4}$) could be established (i.e. cometary water could be formed) in the solar nebula, 
even if the D/H ratio of water ice delivered to the disk was very high ($\sim$10$^{-2}$).

%and C$^{13}$CH + S $\rightarrow$ $^{13}$CCS + H lowers the C$^{13}$CS/$^{13}$CCS ratio.

%We also find that observed $^{12}$C$^{13}$CH/$^{13}$C$^{12}$CX ratio and $^{12}$C$^{13}$CS/$^{13}$C$^{12}$CS ratio are possibly made through the exchange of the $^{13}$C position after formation of molecules. 
%it is found that observed ratios cannot be explained by    
\end{abstract}





\newpage\section{INTRODUCTION}
Water is one of the most important molecules in star and planet forming regions from the perspective 
of the oxygen reservoir and the origin of Earth's ocean.
It is well established that water ice forms on grain surfaces and is already abundant ($\sim$10$^{-4}$ per hydrogen nucleus) in molecular clouds 
at visual extinctions $A_{\rm V} \gtrsim 3$ magnitude \citep[e.g.,][]{whittet93}.
Water has also been detected toward star and planet forming regions, such as prestellar cores \citep{caselli12b}, protostellar envelopes \citep[e.g.,][]{ceccarelli00,boogert08}, and Class II protoplanetary disks \citep[PPDs;][]{terada07,honda09,carr08, hogerheijde11}.
Although there is no doubt that water exists in all the phases of star formation, the evolution of water from molecular clouds to PPDs remains unclear.

D/H ratio can be a useful tool to probe the evolution of water.
Comets are preserved samples of the cold ice-bearing regions in the solar nebula \citep{brownlee03}.
So far, the HDO/H$_2$O abundance ratio has been measured in nine comets; seven from the Oort cloud \citep[$\sim$(4--$10) \times10^{-4}$; e.g.,][]{villanueva09,bockelee-morvan12}, and two from the Kuiper belt \citep[(3--$<\!\!4) \times10^{-4}$;][]{hartogh11,lis13}.
In all sources, the HDO/H$_2$O ratio is higher by about one order of magnitude than the elemental abundance of deuterium in the interstellar medium \citep[$\sim \!\! 1.5 \times10^{-5}$;][]{linsky03}.
On the other hand, to the best of our knowledge, there has been no detection of HDO toward PPDs in the Class II phase except for \citet{ceccarelli05b}.
Although they claimed a 4$\sigma$ detection of HDO vapor toward DM Tau, it has been questioned by \citet{guilloteau06}.

%It is generally believed that water ice (in part) is released to the gas phase inside the sublimation radius ($T_{\rm dust} \gtrsim 150$ K, $r \lesssim$ 100 AU) of protostellar envelopes \citep{aikawa08,crimier10}.
Recently, the HDO/H$_2$O ratio in the gas phase has been measured toward four Class 0--I low-mass protostars, which are summarized in Table \ref{table:hdo}.
We distinguish between the single dish and interferometer observations.
The single dish observations, on the other hand, provide the integrated emission from larger scales.
The interferometer observations provide model-independent estimates of the HDO/H$_2$O ratio in the most inner regions of hot corinos, 
where ice is sublimated \citep[][]{jorgensen10}.
The HDO/H$_2$O ratio derived by using interferometers varies greatly among sources.
The ratios observed towards IRAS 16293-2422 and NGC 1333-IRAS4B are slightly higher than the cometary values \citep{jorgensen10,persson13}, 
whereas the other two sources (NGC 1333-IRAS2A and NGC 1333-IRAS4A) show much higher ratios \citep{taquet13a}.
The difference between the HDO/H$_2$O ratio observed toward comets and protostar envelopes may indicate that water formed 
in the cloud/core phase was reprocessed in the solar nebula.
%This variation of the HDO/H$_2$O ratio may reflect the different physical and chemical conditions, when the ice was formed \citep[e.g.,][]{caselli12a}.

We briefly outline the evolution of HDO/H$_2$O ratio during star and disk formation referring to previous theoretical models.
\citet{visser11} investigated molecular evolution from the collapse of dense cloud cores to the formation of circumstellar disks in two spatial dimensions.
They showed that the majority of water is delivered from the core to the disk as ice.
Although they did not include isotopes, their results indicate that D/H ratio of water delivered to the disk is determined in the parent cloud core.  
In the hot inner regions of disks, where gas temperature is higher than 500 K, isotopic exchange between HDO and H$_2$ occurs, and the HDO/H$_2$O ratio can decrease to 
$\sim$10$^{-5}$ \citep{lecluse94}.
Such reprocessed water could be mixed with unprocessed water via radial mixing and accretion \citep{drouart99,hersant01,yang13}.
The resultant HDO/H$_2$O ratio depends on the thermal history of the disk and efficiency of radial mixing, 
i.e. how much material diffuse from the inner hot region ($>$500 K) to outer radii.
If we assume the disk formation model of \citet{visser11}, the HDO/H$_2$O ratio decreases by a factor of $<$2 on average from the initial cloud value,
since only $\sim$30 \% of the disk mass experiences high temperatures of several 100 K until envelope accretion onto the protostellar system halts.

The primary motivation of this work is to investigate the impact of turbulent mixing in the vertical direction on the HDO/H$_2$O ratio in the Class II PPDs.
Deuterium fractionation in Class II PPDs has been studied numerically by several authors without mixing \citep[e.g.,][]{aikawa01,aikawa02,ceccarelli05,willacy07,willacy09}.
\citet{willacy09} investigated deuterium chemistry in inner disks ($\le 30$ AU) considering the accretion flow.
In the Class II phase, the disk surface is irradiated by stellar ultraviolet (UV) photons and X-rays.
Then, oxygen is mainly in atomic form at the disk surface, while it is in water near the midplane \citep[e.g.,][and references therein]{bergin07}.
They found that the HDO/H$_2$O ratio at the midplane retains their initial value ($\sim \!\! 2 \times10^{-2}$, which is set by their dense cloud core model) for 10$^6$ yr, 
probably because water is not destroyed efficiently in the regions where UV photons and X-rays are attenuated.
However, turbulent mixing could change the situation drastically.
If turbulence exists, water would be transported to the disk atmosphere and destroyed by photoreactions, while atomic oxygen would be transported 
to the disk midplane and reforms water and/or other molecules.
Such destruction and reformation processes are of potential importance for the HDO/H$_2$O ratio in disks;
the HDO/H$_2$O ratio could be significantly changed from that in molecular clouds/cores.

The plan of this paper is as follows.
We describe our physical model and chemical model with vertical mixing in Section 2 and 3, respectively.
In Section 4, we present how the vertical mixing affects water and deuterated water chemistry.
In Section 5, we discuss the effect of initial HDO/H$_2$O ratio, radial mass transport, and dead zone on our results presented in Section \ref{sec:result}.
We also discuss the uncertainties of grain surface chemistry and the implication of our model results for the D/H ratio of cometary water.
We summarize our conclusions in Section 6.

%% In a manner similar to \objectname authors can provide links to dataset
%% hosted at participating data centers via the \dataset{} command.  The
%% second curly bracket argument is printed in the text while the first
%% parentheses argument serves as the valid data set identifier.  Large
%% lists of data set are best provided in a table (see Table 3 for an example).
%% Valid data set identifiers should be obtained from the data center that
%% is currently hosting the data.
%%
%% Note that AASTeX interprets everything between the curly braces in the 
%% macro as regular text, so any special characters, e.g. "#" or "_," must be 
%% preceded by a backslash. Otherwise, you will get a LaTeX error when you 
%% compile your manuscript.  Special characters do not 
%% need to be escaped in the optional, square-bracket argument.   
\section{DISK STRUCTURE} \label{sec:disk_structure}
We adopt the disk model of \citet{nomura05} with the addition of X-ray heating by \citet{nomura07}.
Similar models have been used to study the chemical evolution without deuterium in protoplanetary disks \citep{walsh10,walsh12,heinzeller11}.
Here we briefly describe the disk model.
More details can be found in the original papers \citep{nomura05,nomura07}.

\begin{figure}
\plotone{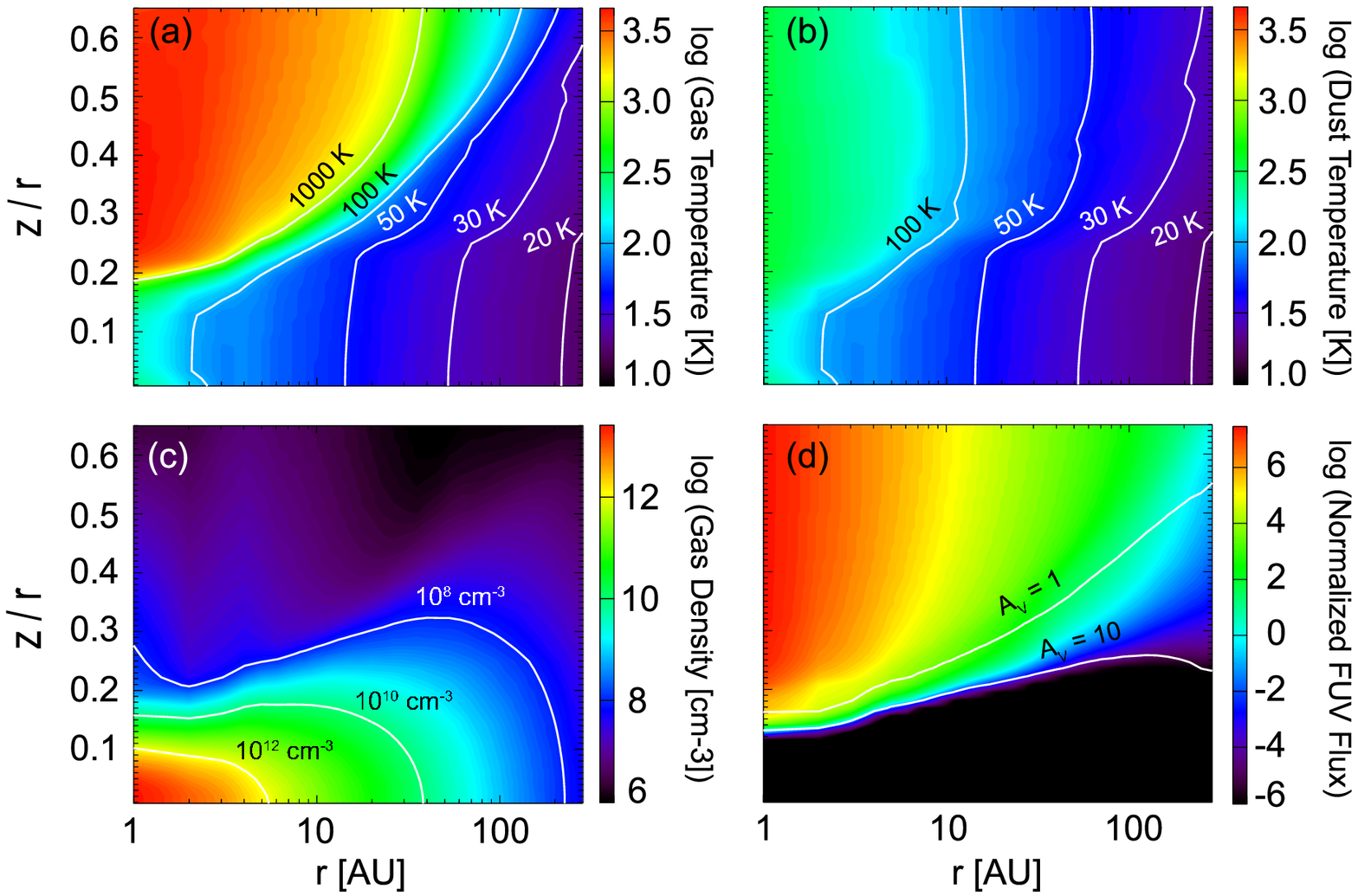}
\caption{Spatial distributions of the (a) gas temperature, (b) dust temperature, (c) number density of gases, and (d) wavelength-integrated FUV flux normalized by Draine field \citep[$1.6\times10^{-3}$ erg cm$^{-2}$ s$^{-1}$; ][]{draine78}. In panel (d), the solid lines indicate the height at which the vertical visual extinction of interstellar radiation field reaches unity and ten.
\label{fig:disk_phys}}
\end{figure}

The model describes the structure of  a steady, axisymmetric Keplerian disk surrounding a typical T Tauri star with the mass $M_{*}=0.5$ $M_{\bigodot}$, 
radius $R_{*}=2$ $R_{\bigodot}$, and effective temperature $T_{*}=4000$ K.
The $\alpha$-disk model \citep{shakura73} is adopted to determine the radial distribution of the surface density, 
assuming the viscous parameter $\alpha = 10^{-2}$, and accretion rate $\dot M=10^{-8}$ $M_{\bigodot}$ yr$^{-1}$.
The gas temperature, dust temperature and density distributions of the disk are calculated self-consistently \citep[see Section 2 of ][for details]{nomura05}.
The stellar UV luminosity is set to be $\sim$10$^{31}$ erg s$^{-1}$ in the range of 6--13.6 eV, while the X-ray luminosity is set to be $\sim$10$^{30}$ erg s$^{-1}$ 
in the range of 0.1--10 keV based on the spectrum observations of TW Hydra \citep[e.g.,][]{herczeg02,kastner02}.
In the model, the dust-to-gas mass ratio is assumed to be 10$^{-2}$ with the dust-size distribution model for dense clouds \citep{weingartner01}.
The resultant (a) gas temperature, (b) dust temperature, (c) number density of gases, and 
(d) wavelength-integrated FUV flux normalized by Draine field \citep{draine78} are shown in Figure \ref{fig:disk_phys}.

\section{CHEMICAL MODEL}
We use a two-phase (gas and grain surface) model to compute disk chemistry \citep{hasegawa92}.
Our chemical reaction network is based on \citet{aikawa12}, who investigated deuterium fractionation 
in the star formation from a prestellar core to a protostellar core. 
They extended the network of \citet{garrod06} to include multi deuterated species and isotope exchange reactions from \citet{millar89} and \citet{roberts04}.
More detailed explanation can be found in \citet{aikawa12}.
To reduce the computational time, we exclude the species with chlorine and phosphorus, molecules with more than four carbon atoms, 
and relevant reactions.
We also make several modifications to make the network model applicable to disk chemistry, which are described below.

In the regions where the gas temperature exceeds 100 K, we use the reaction set based on \citet{harada10} instead of those of \citet{garrod06} for the neutral-neutral 
and neutral-ion reactions.
The network of \citet{harada10} was designed for modeling high-temperature gas phase chemistry (100--800 K).
This switching method of the reaction sets was also used in \citet{furuya12} to simulate the chemical evolution in collapsing molecular cloud cores.
For high temperature reactions of water and its isotopologues, we adopt the rates in \citet{talukdar96}, \citet{bergin99}, 
and the UMIST database \citep{woodall07}.
They are summarized in Table \ref{table:react}.
We include some three-body association reactions and the reverse reactions (i.e. collisional dissociation) based on \citet{willacy98} and \citet{furuya12}.
We also modify or include the following reactions, which are important for disk chemistry: 
(i) X-ray chemistry (Section \ref{sec:ionizaion} and \ref{sec:photochem1}), (ii) photoreactions by stellar and interstellar UV (Section \ref{sec:photochem1}),
 (iii) chemistry of vibrationally excited H$_2$ and its isotopologues (Section \ref{sec:vibh2}).
The treatment of gas-grain interactions and grain surface reactions is described in Section \ref{sec:gas-grain}.
In total, our reaction network consists of 726 gaseous species, 303 grain-surface species, dust grains with three different charge states ($\pm$1, 0), 
and 34,997 (38,162) reactions for gas temperatures $\leq$100 K ($>$100 K).

\subsection{Ionization Rate}\label{sec:ionizaion}
Interstellar cosmic-rays and X-rays from a central star are the main ionization sources of protoplanetary disks.
The cosmic-ray ionization rate of hydrogen molecules is expressed as
\begin{equation}
\xi_{{\rm CR}}(r, z) = \frac{\xi^0_{\rm CR}}{2}\left[\exp\left(-\frac{\Sigma_1(r, z)}{96\,{\rm g\,cm^{-2}}}\right) + \exp\left(-\frac{\Sigma_2(r, z)}{96\,{\rm g\,cm^{-2}}}\right)\right],
\end{equation}
where $\Sigma_1(r, z)$ and $\Sigma_2(r, z)$ are the vertical column density of gas measured from the upper and lower boundary of the disk to the height $z$, respectively \citep[e.g.,][]{semenov04}.
Unattenuated cosmic-ray ionization rate, $\xi^0_{\rm CR}$, is set to be $5\times10^{-17}$ s$^{-1}$ \citep{dalgarno06}.
The attenuation length of cosmic-ray is 96 g cm$^{-2}$ \citep{umebayashi81}, which is larger than our disk column density only 
at $r \lesssim 1$ AU.
We also take into account the ionization by short-lived radio active isotopes with the rate of $1\times10^{-18}$ s$^{-1}$ \citep{umebayashi09}.

The primary ionization rate of species $i$ by X-rays is given by 
\begin{equation}
\xi_{{\rm prim}, i}(r, z) = \int_{}^{}F_{\rm X}(E, r, z)\sigma_{{\rm pi}, i}(E)dE,
\end{equation}
where $F_{\rm X}$ is the photon number flux from the central star.
The photoionization cross sections, $\sigma_{{\rm pi}, i}$, are taken from \citet{verner95}.
We consider the primary ionization of atoms, singly ions, and diatomic molecules.
Considering the Auger effect, it is assumed that ionization of atoms leads to doubly charged ions, while ionization of diatomic 
molecules produces a pair of singly charged ions, following \citet{maloney96}.
The cross sections of the molecules are calculated by adding the atomic cross sections.
We use the same cross sections for deuterated species as normal isotopes.
Electron recombination and charge transfer reactions for doubly ionized species are taken from \citet{stauber05}.
We neglect doubly ionized states of Na and Mg, since the elemental abundances of Na ($2.25\times10^{-9}$) and Mg ($1.09\times10^{-10}$) are low in our model.
We also neglect doubly ionized state of Si, since its abundance is much lower than that of Si$^{+}$ in PPDs \citep[e.g.,][]{adamkovics11}.

Part of the kinetic energy of photoelectrons produced by primary ionization is lost by ionizations of H, H$_2$, and He.
These secondary ionizations are far more important for the ionization degree of gas than the primary ionization.
Assuming that the kinetic energy of Auger electrons is equal to the photoionisation threshold energy \citep{voit91}, the overall 
secondary ionization rate per hydrogen nucleus is 
\begin{equation}
\xi_{\rm sec}(r, z) = \sum_{i}\int_{}^{}F_{\rm X}\sigma_{\rm eff}(E)\frac{E}{W_i(E)}dE,
\end{equation}
where $\sigma_{\rm eff}$ ($=\sum_{i}^{} x^{\rm cosmic}_{i}\sigma_{{\rm pi}, i}$) is the effective photoionization cross section 
per hydrogen nucleus evaluated assuming a cosmic abundance for elements ($x^{\rm cosmic}_{i}$).
\citet{dalgarno99} derived the mean energy required per secondary ionization of species $i$, $W_i$, for the H-H$_2$-He mixtures as a function of 
the ionization degree and the abundance ratio of H and H$_2$.
We use Equations (9), (10), (13) and (15) of \citet{dalgarno99} with the parameter applicable to a 1 keV electron to calculate $W$ in our models \citep[e.g.,][]{meijerink05}.
Although secondary ionizations of other species have no significance for the ionization degree of gas, it could be important for chemistry \citep{maloney96}.
The secondary ionization rate of other species $i$ per hydrogen nucleus is estimated by
\begin{equation}
\xi_{{\rm sec}, i} = \xi_{\rm sec}\frac{x_i\sigma_{{\rm ei}, i}}{x_{\rm H}\sigma_{\rm ei, H} + x_{\rm H_2}\sigma_{\rm ei, H_2} + x_{\rm He}\sigma_{\rm ei, He}},\label{eq:seci}
\end{equation}
where $x_{i}$ represent the abundance of species $i$ relative to hydrogen nuclei.
The electron impact cross sections, $\sigma_{\rm ei}$, are taken from the NIST database (http://www.nist.gov/pml/data/ionization/) and \citet{lennon88}.
We include secondary ionization of CO, singly charged atomic ions, and atoms except for H and He in our network using Equation (\ref{eq:seci}).
In our disk models, the secondary ionization by X-rays is the dominant ionization source except for the midplane (e.g., $N_{\rm H} \gtrsim 10^{24}$ cm$^{-2}$ at 5 AU, 
where $N_{\rm H}$ is the vertical column density of hydrogen nucleus from the disk surface.), where cosmic ray ionization is dominant.
%We use $\xi_{{\rm T}, i}$ to calculate rates of reactions for species $i$ with cosmic-ray particles, while use $\xi_{\rm T, H_2}$ to calculate cosmic-ray-induced photo reactions.

\subsection{Photochemistry} \label{sec:photochem1}
Protoplanetary disks are irradiated by UV photons from interstellar radiation and central stars.
While the former has a continuous spectral energy distribution, the latter is the combination of continuum and Ly$\alpha$ radiation.
Furthermore, cosmic-rays and X-rays induce UV photons through energetic photoelectrons.
The UV spectrum at each point in the disk is given as a combination of them.
Because of the different origins and wavelength dependence, photochemistry by these photons is considered separately.
In this subsection, rate coefficients (s$^{-1}$) are denoted by $k$, while photorates (cm$^{-3}$ s$^{-1}$) are denoted by $R$.

\subsubsection{UV Continuum} \label{sec:photochem2}
Spectral energy distribution of UV continuum is given at each point in the disk (Section \ref{sec:disk_structure}).
To calculate photorates, we convolve the UV spectrum with absorption cross sections.
Referring to \citet{van dishoeck88}, the rate coefficients of photodissociation and photoionization in the gas phase (s$^{-1}$) are calculated as
\begin{equation}
k_{{\rm phg},i}(r, z) = \int_{}^{} 4\pi J(\lambda, r, z) \sigma_i(\lambda) d\lambda, \label{eq:kphg}
\end{equation}
%\begin{equation}
%k_{\rm ph, g}^{\rm line}(R, z) = \frac{\pi e^2}{mc^2}\lambda_{\rm ul}^2f_{\rm ul}\eta_{\rm u}I_{\lambda_{\rm ul}}(R, z), \label{eq:kline}
%\end{equation}
where $J(\lambda)$ is the the mean intensity of the radiation field measured in photons cm$^{-2}$ s$^{-1}$ \AA $^{-1}$ str$^{-1}$ at the wavelength of $\lambda$.
The cross sections, $\sigma(\lambda)$, are given by \citet{van dishoeck88}, and updated by \citet{jansen95a,jansen95b} and \citet{van dishoeck06}.
The data is downloadable from http://www.strw.leidenuniv.nl/\~{}ewine/photo/.
For deuterated species, we use the same cross sections as normal species.
For species which are not in the database, we use the cross sections of a similar type of species \citep[e.g.,][]{van zadelhoff03}.

H$_2$, HD, and CO are dissociated by absorption lines.
Because of their high abundances, self-shielding and mutual shielding of HD and CO by H$_2$ are important.
The self-shielding factor of H$_2$ is given as a function of H$_2$ column density referring to Equation (37) of \citet{draine98}.
Since the self-shielding behavior of HD is almost identical to that of H$_2$ \citep{wolcott11}, we also use Equation (37) of \citet{draine98} for self-shielding factor of HD 
by replacing H$_2$ column density with that of HD.
The shielding factor of HD by H$_2$ is given by Equation (12) of \citet{wolcott11}.
%, and we assume that its upper value (i.e. when the overlapped HD lines are totally shielded by H$_2$) is 1/3 referring to \citet{barsuhn77}.  
For CO, the self-shielding and mutual shielding factors are calculated following \citet{lee96}.

Then, the photodissociation rate coefficient of CO, for example, is calculated as follows:
\begin{equation}
k_{\rm phg}(r, z) = k^{0}_{{\rm phg}, r}f_{\rm sh}(N_{r, {\rm CO}}, N_{r, {\rm H_2}}) + k^{0}_{{\rm phg}, z}f_{\rm sh}(N_{z, {\rm CO}}, N_{z, {\rm H_2}}),
\end{equation}
where $k^{0}_{{\rm phg}, r}$ and $k^{0}_{{\rm phg}, z}$ are non-shielding photodissociation rates of CO, which are calculated at each point using Equation (\ref{eq:kphg}), by the UV radiation in the radial and vertical directions, respectively.
The shielding factor, $f_{\rm sh}$, is a function of the column densities of CO and H$_2$, $N_{{\rm CO}}$ and $N_{{\rm H_2}}$, along the ray.
In the vertical direction, we obtain the shielding factor at each vertical position $z$ by integrating the number density of CO and H$_2$ at $>\!\!z$,
\begin{equation}
N_{z, i}(r, z) = \int^{\infty}_{z}n_{i}dz',
\end{equation}
where $n_{i}$ represents the number density of species $i$.
In the radial direction, on the other hand, molecular column densities are assumed to be the overall column density along the ray multiplied by the local molecular abundances:
\begin{equation}
N_{r, i}(r, z) = x_{i}(r, z)\int^{r}_{R_*}n_{\rm H}dr', \label{eq:sigmar}
\end{equation}
where $n_{\rm H}$ represent the number density of hydrogen nuclei.
Equation (\ref{eq:sigmar}) could overestimate the column densities (i.e. shielding factors), since the radiation field decays and the abundances of CO and H$_2$ would increase along the ray.

Grain surface species are also photodissociated and photodesorbed by UV photons \citep[e.g.,][]{gerakines96,westley95}.
The photoabsorption cross sections of H$_2$O ice and CO$_2$ ice are taken from \citet{mason06}.
It is worth noting that the cross sections of ices are different from those in the gas phase \citep[see Figure 1 of][ for water]{andersson08}. 
For the other species, we use the same cross section for corresponding gaseous species, because data is not available in the literature.
We confirmed that our results are not sensitive to this assumption.
We performed a calulation using the ten times smaller cross sections for ice molecules except for water and CO$_2$ ices, 
and found that the results are not significantly different from those in Section 4.
%We use the same cross sections and oscillator strengths for photodissociation rates of grain surface species as those of corresponding gaseous species, since most species are weakly bound on grain surfaces \citep{ruffle01}.
The photodissociation rates on grain surfaces (cm$^{-3}$ s$^{-1}$) are calculated as
\begin{align}
%k_{ph, d}(R, z) = 4\pi a^2\Sigma_{site}n_d k_{ph, g}(R, z)/n_{ice},
%k_{\rm ph, d}(R, z) = k_{\rm ph, g}f_{\rm ice}N_{\rm p}/N_{\rm layer}, \label{eq:kphd}
%R_{\rm ph, d}(r, z) = n_{\rm d}\theta_iN_{\rm site}\int_{0}^{\infty} 4\pi I_{\lambda}P_{\lambda}\sigma_{\rm site} d\lambda,
R_{{\rm phs}, i}(r, z) & = \pi a^2n_{\rm d}\theta_i \int_{}^{} 4\pi J(\lambda, r, z)P_i(\lambda) d\lambda, \label{eq:rphs}\\
\theta_i & = \frac{n_i}{{\rm max}(n_{\rm ice}, n_{\rm d}N_{\rm site})}, \label{eq:thetai}\\
P_i(\lambda) & = {\rm min}\left(1, \frac{N_{\rm p}\sigma_i(\lambda)}{\sigma_{\rm site}}\right), \label{eq:pi}
%N_p & = {\rm max}[1, {\rm min}(N_{\rm layer}, 5)], \label{eq:pi}
\end{align}
where $n_{\rm d}$, $n_{\rm ice}$, $\theta_i$, and, $N_{\rm site}$ are the number density of the grain, the total number density of grain surface species, the coverage of species $i$ on a grain, 
and the number of adsorption sites on a grain ($N_{\rm site} \sim 10^6$), respectively.
The grain radius, $a$, is set to be 0.1 $\mu$m in our chemical model for simplicity.
The ratio of the geometrical and absorption cross sections of dust grains for FUV photons is assumed to be unity \citep[e.g.,][]{tielens05}. 
$P_i(\lambda)$ and $\sigma_{\rm site}$ are the probability for an incident photon with the wavelength of $\lambda$ on a site to be absorbed 
by an adsorbed species and the site size ($4\pi a^2/N_{\rm site}$), respectively.

Every photoabsorption does not necessarily result in photodissociation.
\citet{andersson06} and \citet{andersson08} carried out molecular dynamics simulations to study photodissociation of water ice in the top six monolayers.
They found that photolytic effect is more significant in the uppermost few monolayers; 
although photons can penetrate into deeper layers and dissociate embedded molecules, the photoproducts recombine with a higher probability than  in uppermost layers.
In this study, we assume that the upper most one monolayer can be dissociated as the outcome of photoabsorption, and that the photoproducts immediately recombine in deeper layers.
For considering surface roughness, parameter $N_{\rm p}$ in Equation (\ref{eq:rphs}) is set to be two monolayers for all species.
We discuss the effect of this assumptions on our results in Section \ref{sec:uncertainty}.

%If $N_{\rm p}$ is set to be $\gtrsim$2, however, we would overestimate the photolytic effect, since our model does not consider the layered structure of ice mantle.
%, and $N_{\rm p}=2$ is a correction for the fact that the photodissociation is efficient only in the first few monolayers  . 

In this study, we adopt a two-phase model, in which a layered structure of ice mantles is not considered and the bulk of ices is chemically active.
More sophisticated ice mantle modeling, which distinguish a chemically active surface layer from an inactive inert bulk, has been developed \citep[][]{hasegawa93,taquet13b}.
%But they are mostly performed in a simple pseudo-time dependent model (i.e. one box model with constant density and temperature).
Since the calculation of the rate equations combined with turbulent mixing is time consuming, we postpone the inclusion of such layered ice structure to future works. 
The combination of the two-phase model and photodissociation of HDO ice could, however, introduce artificial decrease of the D/H ratio of water ice as follows.
Photodissociation of HDO ice has two branches, producing OH ice or OD ice.
While OH ice is able to cycle back to HDO ice via the grain surface reactions, e.g., OH$_{\rm ice}$ + D$_{\rm ice}$, it is also converted to H$_2$O ice.
If photodissociation of ices is efficient only in the uppermost several layers, this cycle could decrease the D/H ratio of water ice on the surface, 
but not the ratio in the bulk water ice.
In the two phase model, on the other hand, this cycle could decrease the D/H ratio in the bulk water ice; 
we do not distinguish between D/H ratios in the surface layers and the deeper layers, although we calculate the photodissociation rate of ice assuming 
that only the uppermost one layer can be dissociated.
To keep the consistency in our model, we switch off the branch to produce OH ice in photodissociation of HDO ice.

The photodesorption rates are calculated as \citep[e.g.,][]{visser11}
\begin{equation}
%k_{pd}(R, z) = \pi a^2n_dF_{\rm UV}(R, z)Y_{pd}/n_{ice}.
R_{{\rm pd}, i}(r, z) = \pi a^2 n_{\rm d}\theta_i F_{\rm FUV}(r, z)Y_i, \label{eq:pd}
\end{equation}
where $F_{\rm FUV}$ and $Y_i$ are the FUV photon number flux integrated in the range of 912--2000 {\AA} and the photodesorption yield per incident FUV photon, respectively.
Some experiments into the photodesorption of UV-irradiated ices have been conducted \citep[e.g.,][]{westley95,oberg07}.
Recently, \citet{oberg09a,oberg09b} measured the yield of CO$_2$ and H$_2$O ices by photons with the wavelength of 1180--1770 {\AA}.
\citet{fayolle11,fayolle13} measured wavelength-dependent photodesorption yields of CO, O$_2$, and N$_2$ ices, and calculated the yields adequate for astrophysical environments. 
For the above species, we use the yields obtained by the experimental works.
The yield for the other species is assumed to be 10$^{-3}$.
We confirmed that the results presented in this paper does not change, if we use the yield of 10$^{-4}$ for species without laboratory data.

%, and its lower limit is set to 10$^4$ cm$^{-2}$ s$^{-1}$ to account for cosmic-ray induced photons \citep{shen04}.
%The photodesorption yield per incident photon, $Y$, is experimentally determined for some species, and they are typically $\sim$10$^{-3}$ except for N$_2$ ice, $\sim$10$^{-4}$.

\subsubsection{Ly$\alpha$ Photons}%\label{sec:photochem2}
Ly$\alpha$ radiation could dominate the UV field in protoplanetary disks \citep{herczeg04}.
We include photoreactions by Ly$\alpha$ photons for a limited number of gaseous species, 
for which photodissociation cross section at Ly$\alpha$ wavelength ($\sim$1216 {\AA}) is available in \citet{van dishoeck06}, and corresponding species in the solid phase.
The cross sections for H$_2$O ice and CO$_2$ ice are taken from \citet{mason06}, while we use the cross section of the gaseous species for the other ice-mantle species.
Considering that the photodissociation of ice-mantle species causes photodesorption \citep{andersson08}, the photodesorption yields per incident Ly$\alpha$ photon, 
$Y_{\rm Ly\alpha}$, are estimated by
\begin{equation}
\frac{Y_{{\rm Ly\alpha}, i}}{Y_i} = \frac{\sigma_i(\lambda_{\rm Ly\alpha})}{\tilde{\sigma_i}},
\end{equation}
where $\sigma_i(\lambda_{\rm Ly\alpha})$ and $\tilde{\sigma_i}$ are the photodissociation cross section at Ly$\alpha$ wavelength and 
the averaged cross section in the wavelength of 1180--1770 {\AA}, respectively.

Recently, \citet{bethell11} showed that the Ly$\alpha$/FUV-continuum photon density ratio is smaller in the regions near the photodissociation layers 
for hydrogen ($N_{\rm H} \lesssim 5\times10^{20}\; {\rm cm^{-2}}$ at 1 AU if the dust is well mixed with gas), 
but larger in the deeper regions than the intrinsic ratio in the incident FUV radiation from the central star, due to resonant scattering of Ly$\alpha$ by atomic hydrogen.
%They also showed that such trend is stronger in the inner disk, but the threshold vertical column density does not strongly depend on the radius.
Since the intensity of Ly$\alpha$ radiation is calculated without considering resonant scattering in the physical model used here,
it would be overestimated in the upper layers of the disk, and underestimated near the midplane.

%The error is, however, less than one order of magnitude, as long as dust is well mixed with gas.
%It also should be noted that we could overestimate the FUV continuum, since dust scattering is treated as the isotropic process in our disk models.
%\citet{fogel11}, however, found that including photoreactions by Ly$\alpha$ photons does not significantly change the disk chemistry as long as dust grains are well mixed with gas.

%lyalpha '¢'ê'闝—R: gas 'Æ solid 'Åsigma'ªˆá'¤'Æbalance•Ï'í'é‰Â"\«

\subsubsection{Cosmic-ray and X-ray Induced UV Photons}%\label{sec:photochem3}
UV photons are generated by the decay of electronically excited hydrogen molecules and atoms produced by collisions with energetic photoelectrons following cosmic-ray ionization.
In dense regions where the stellar and interstellar UV photons are heavily attenuated, cosmic-ray induced photons dominate the UV fields \citep{prasad83}.
X-ray ionization also induces UV photons in a similar manner.
Our model includes photochemistry by cosmic-ray and X-ray induced UV photons both in the gas phase and on grain surfaces.
The rates are described by the local competition of photon absorption by gaseous species and dust grains.
Referring to \citet{gredel89}, the rate coefficients of photodissociation and photoionization of species $i$ in the gas phase (s$^{-1}$) are expressed as
\begin{align}
k_{{\rm crphg}, i}(r, z) & = x_{\rm H_2}\xi_{\rm T}\int_{}^{} \frac{\sigma_{i}(\lambda)}{\sigma_{\rm tot}}\psi(\lambda)d\lambda, \label{eq:crphg}\\
\sigma_{\rm tot} & = x_{\rm d}\pi a^2, \label{eq:sigmat}\\
\xi_{\rm T} & = \xi_{\rm CR}+\xi_{\rm sec},
\end{align}
%\begin{equation}
%p_{\rm M}=\int_{0}^{\infty} \frac{\sigma_{\nu}f_{\nu}}{\sigma_{\rm g}} d\nu,
%\end{equation}
%,where $\xi_{\rm CR}$ and $\xi$ are the cosmic-ray ionization rate and the total ionization rate, respectively (see Section \ref{sec:ionizaion}).
%The flux of cosmic-ray induced photons, $F_{\rm CRUV}$, is assumed to be $10^4$ cm$^{-2}$ s$^{-1}$ \citep{shen04}.
%The term $\int p_{\nu} \sigma_{\nu}d\nu$, which depends on the photodissociation or photoionization cross sections of each species, was originally calculated by \citet{gredel89}, and is available in the Ohio State University network.
where $\psi(\lambda)$, and $x_{\rm d}$ are the number of photons with the wavelength of $\lambda$ produced per ionization, and the relative abundance of 
dust grains to hydrogen nucleus ($\sim \!\! 2\times10^{-12}$), respectively.
Again, the ratio of the geometrical and absorption cross sections of dust grains for FUV photons is assumed to be unity \citep[e.g.,][]{tielens05}.
For simplicity, we ignore photons produced by the de-excitation of atomic hydrogen and assume $x_{\rm H_2}=0.5$ in Equation (\ref{eq:crphg}).
It is an acceptable assumption, since the regions dominated by atomic hydrogen do not contain many molecules \citep{aikawa99}.
We also ignore the gas opacity in Equation (\ref{eq:sigmat}).
If all oxygen is in gaseous water, the latter assumption overestimates the rate coefficients by a factor of about two at maximum, 
since $x_{\rm H_2O}\tilde{\sigma}_{\rm H_2O}$ is $\sim \!\! 7 \times 10^{-22}$ cm$^{-2}$ and $x_{\rm d}\pi a^2$ is $\sim \!\! 6 \times 10^{-22}$ cm$^{-2}$.
However, it is also acceptable, since induced UV photons are more important than stellar UV photons only near the midplane, 
and dust temperatures near the midplane in our disk models are lower than the sublimation temperature of water ice except for $r\lesssim1$ AU.
%since most abundant gaseous species is CO near the midplane where induced photons are more important than stellar UV photons, and it only absorbed photons at specific wavelength. 

The rates (cm$^{-3}$ s$^{-1}$) of photodissociation on grain surfaces and photodesorption are calculated by 
\begin{align}
R_{{\rm crphs}, i}(r, z) &= n_{\rm H_2}\xi_{\rm T} \theta_i \int_{}^{} \psi(\lambda)P_i(\lambda) d\lambda,\\
R_{{\rm crpd}, i}(r, z) &= n_{\rm H_2}\xi_{\rm T} \theta_i Y_i\int_{}^{}\psi(\lambda)d\lambda.
\end{align}
%The photodesorption rates by induced UV photons are calculated in a similar manner as Eq. \ref{eq:pd}, replacing $F_{\rm UV}$ to $\varepsilon F_{\rm CRUV}$.

\subsubsection{Photoreaction Timescales in the Disk}
\begin{figure}
\epsscale{0.8}
\plotone{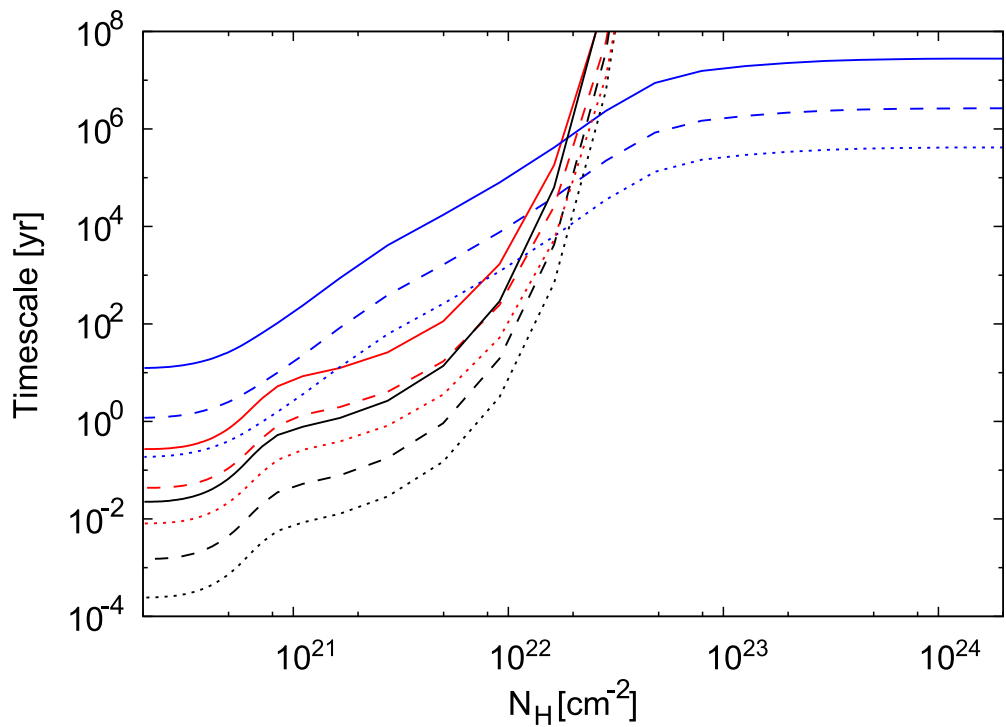}
\caption{Timescales of photoreactions of water ice and water vapor as functions of vertical column density in our disk model at $r=10$ AU. The solid, dashed, and dotted lines indicate timescales of photodesorption, photodissociation on grain surfaces, and photodissociation in the gas phase, respectively. The black, red, and blue lines indicate photoreactions by Ly$\alpha$ photons, FUV continuum, and X-ray and cosmic-ray induced photons, respectively.\label{fig:photorate}}
\end{figure}

Figure \ref{fig:photorate} shows the timescales of photoreactions of water ice and water vapor as a function of vertical column density in our disk model at $r=10$ AU.
Here, the photoreaction timescales of water ice are defined as the timescales in which one monolayer of pure water ice is photodissociated (photodesorbed), 
$n_{\rm d}N_{\rm site}/R(\theta_{\rm H_2O_{ice}}=1)$, while those of water vapor are defined as an inverse of the rate coefficients, $k^{-1}$.
It is clear that the photoreactions by Ly$\alpha$ photons are most efficient in the disk atmosphere, 
while the photoreactions by X-ray and cosmic-ray induced UV photons are the most efficient at $N_{\rm H} \gtrsim 2 \times 10^{22}$ cm$^{-2}$ at 10 AU.
It should be noted, however, that these timescales are larger 
than the typical age of T Tauri stars ($\sim$10$^6$ yr) at $N_{\rm H} \gtrsim 10^{23}$ cm$^{-2}$.

%\begin{equation}
%R_{\rm crpd} = n_{\rm d}\theta_iN_{\rm site}\sigma_{\rm site}F_{\rm CR, UV}\left(\frac{\xi}{\xi_{\rm CR}}\right)Y_{\rm pd}.
%\end{equation}
\subsection{Vibrationally Excited Hydrogen Molecule}\label{sec:vibh2}
Hydrogen molecules can be electronically excited by absorbing FUV photons.
Subsequent fluorescence leads to dissociation or to the population of various rotation-vibration levels in the ground electronic state.
The branching ratio of the former and latter are 10 \% and 90 \%, respectively \citep{black76}.
Following \citet{tielens85}, we consider only two states of H$_2$, the ground vibrational state and a single vibrationally excited state (H$_2^{\ast}$). 
They are treated as different species in our chemical model.
The effective quantum number of this pseudolevel is $v=6$, and the effective energy is 2.6 eV \citep[$\sim \!\! 3\times10^4$ K,][]{london78}.
The FUV pumping rate is set to be 9 times the photodissociation rate of H$_2$.
H$_2^{\ast}$ can be destroyed by photodissociation, radiative decay, collisional de-excitation, or collisional dissociation.
These rates are calculated following \citet{tielens85}. 
As an inverse reaction of the collisional de-excitation, we consider collisional excitation following \citet{woitke09a}.

H$_2^{\ast}$ reacts with other species using internal energy to overcome activation barriers.
It should be noted, however, that whether the internal energy of H$_2^{\ast}$ is effectively used or 
not is quite specific for each reaction and difficult to predict as discussed by \citet{agundez10}. 
Therefore, it would be problematic to simply assume that all the internal energy of H$_2^{\ast}$ is used to overcome activation barriers in any reactions.
To be conservative, we include a limiting number of reactions between H$_2^{\ast}$ and other species (He$^+$, C$^+$, O, OH, and CN), 
rate coefficients of which are experimentally or/and theoretically determined \citep[see Table 1 of ][]{agundez10}. 
Vibrationally excited HD and D$_2$ are treated in the same way as H$_2^{*}$.

\subsection{Gas Grain Interactions and Surface Reactions}\label{sec:gas-grain}
Calculations of gas-grain interactions and grain-surface reactions except for H$_2$ formation are performed in a similar way to \citet{garrod06} and \citet{furuya12}.
A brief summary is provided below.

The sticking probability of neutral species onto dust grains is assumed to be unity except for atomic hydrogen and deuterium; 
their sticking probabilities are calculated as a function of gas and dust temperatures following \citet{hollenbach79}.
Interactions between ions and dust grains are calculated in the same way as \citet{furuya12}.
We adopt the same desorption energies ($E_{\rm des}$) of atoms and molecules as \citet{garrod06}, in which grain-surfaces are assumed to be covered with non-porous 
water ice, and thus only physisorption sites are considered, unless stated otherwise.
For deuterated species, we use the same desorption energies as normal species except for atomic deuterium, whose adsorption energy is set 21 K higher than that of atomic hydrogen, 
following \citet{caselli02}.
$E_{\rm des}$ of selected species are listed in Table \ref{table:edes}.
In addition to the thermal desorption and photodesorption (see Section \ref{sec:photochem1}), 
we consider sublimation via stochastic heating by cosmic-ray \citep{hasegawa93} and X-ray, and the chemical desorption.
We assume that roughly 1\% of species formed by surface reactions are desorbed following \citet{garrod07}.

Grain surface reactions are assumed to occur by the Langmuir-Hinshelwood (LH) mechanism between physisorbed species; adsorbed species diffuse by thermal hopping and react 
with each other when they meet \citep[e.g.,][]{hasegawa92}.
The energy barrier against diffusion is set to be a half of the desorption energy.
If surface reactions have activation energy barriers, they are overcome thermally, or via quantum tunneling, whichever is faster \citep{garrod06}.
We use the modified rate method \citep{caselli98} for surface reactions which include atomic hydrogen or deuterium as reactants.

In our disk models, H$_2$ formation by association of physisorbed H atoms does not efficiently occur by the LH mechanism, 
since the dust temperatures are $\gtrsim$20 K except for the midplane at $r \gtrsim 250$ AU \citep[e.g.,][]{hollenbach71}.
\citet{cazaux04,cazaux10} and \citet{cazaux08} studied formation of H$_2$ and its isotopologues on $bare$ grain surfaces, considering both physisorbed and chemisorbed sites 
and both the LH and Eley-Rideal mechanisms.
They showed that H$_2$ and HD can be formed with the efficiency of several tens of percent until quite high temperatures (several hundreds K).
Since the abundance of water in disk atmospheres is highly dependent on the abundance of H$_2$ \citep[e.g.,][]{glassgold09}, H$_2$ formation 
on bare grain surfaces should be considered.
The precise efficiency is, however, highly dependent on the width and height of the barrier between the adsorption sites, 
which is not well constrained; 
we adopt a constant efficiency of 0.2 for H$_2$ and HD independent of dust temperature.
We ignore D$_2$ formation on bare grain surfaces, since it is not as efficient as that of H$_2$ and HD \citep{cazaux08}. 
Considering the accretion rate of a pair of atomic hydrogen onto bare grain surfaces, the formation rate of H$_2$ is expressed as
\begin{equation}
R_{\rm H_2} = \frac{1}{2}\pi a^2n_{\rm d}n_{\rm H}S\varepsilon_{\rm H_2}(1-\theta_{\rm ice})\sqrt{\frac{8kT_{\rm g}}{\pi m_{\rm H}}},
\end{equation}
where $\varepsilon_{\rm H_2}$ is the formation efficiency.
The coverage of ice mantle species, $\theta_{\rm ice}$, is given by substituting $n_{\rm ice}$ into $n_i$ in Equation (\ref{eq:thetai}).
The rate is set to zero when dust grains are covered by more than one monolayer of ice (i.e. $n_{\rm ice} \ge n_{\rm d}N_{\rm site}$).

\subsection{Initial Abundances for Disk Chemistry \label{sec:initial}}
%\citet{willacy07,willacy09} used the output abundances of a molecular cloud model as the input abundances of their disk models.
%They show that  D/H ratios for some species affect disk chemistry, but D/H ratios for some species retain their input values for 10$^6$ yr. It indicates that the initial abundances of disks are important for deuterium chemistry in disks.

%Deuterium fractionation is likely to occur in low temperature conditions, such as molecular cloud cores \citep[e.g.][]{roberts04}.
%After the protostar and disk formation, materials in the surrounding envelope are heated by radiation from the central star, and experience shock heating when they accrete to the disk.
%\citet{aikawa12} investigated deuterium chemistry from a molecular cloud core to a protostellar core adopting spherically symmetric radiation hydrodynamic simulations \citep{masunaga00}.
%They found that fractionation in ion molecules 

%In such warm or high temperature conditions, the deuterium fractionation is possibly relaxed.
%To gain the initial abundances of our disk model, we revisit the calculation of \citet{aikawa12} using our network. 
%in the parcel which reaches $r=15$ AU at the protostellar age of $9.3\times10^4$ yrs.

We adopt the so-called low metal values as the elemental abundances \citep[see Table 1 of][]{aikawa01}. 
The elemental abundance of deuterium is set to be $1.5 \times 10^{-5}$ \citep{linsky03}.
It is assumed that hydrogen and deuterium are initially in H$_2$ and HD, respectively.
The heavy elements are assumed to be initially in atomic or ionic form, corresponding to their ionization energy.
We integrate the rate equation, using our chemical network model, in the collapsing core model of \citet{aikawa12}, 
in which fluid parcels are traced from the prestellar core to the protostellar core of age $9.3\times10^4$ yr.
Because the abundances are mostly constant at $r \lesssim 100$ AU in this protostellar core model, 
we adopt the molecular abundances at $r=60$ AU as the initial abundance of our disk model (Table \ref{table:initial}).
%We integrate the initial abundances for 10$^6$ yr under the typical dense cloud core conditions; $n_{\rm H}=2\times10^4$ cm$^{-3}$, $T=10$ K, $A_{\rm V} = 10$ mag.

%The updated abundances are further processed to make the input abundances of our disk models.
%We determine what fraction is initially in the gas phase and on grain surfaces from the balance between the freeze-out timescale, $R_{\rm acc}^{-1}$, and the thermal desorption timescale, $R_{\rm td}^{-1}$, for each species at each disk point:
%\begin{equation}
%f = R_{\rm acc}/R_{\rm td}, \;\;\;\;\;\; n_{i, \rm{gas}}^0 = \frac{n_{i, {\rm gas}}^{\rm cc}+n_{i, {\rm ice}}^{\rm cc}}{1+f}, \;\;\;\;\;\; n_{i, \rm{ice}}^0 = \frac{n_{i, {\rm gas}}^{\rm cc}+n_{i, {\rm ice}}^{\rm cc}}{1+f}f, 
%\end{equation}
%where $n^0$ and $n^{\rm dc}$ are the input abundances for our disk models and the output abundances of the core model, respectively.

\subsection{Turbulent Mixing}
We compute molecular evolution in the protoplanetary disk with vertical mixing, assuming that the origin of 
the disk turbulence is the magnetorotational instability \citep[MRI;][]{balbus91}.
We consider the inhomogeneous one-dimensional diffusion equations at a specific radius \citep[e.g.,][]{xie95,willacy06},
\begin{align}
\frac{\partial n_i}{\partial t} + \frac{\partial \phi_i}{\partial z} = P_i - L_i, \label{eq:basic}\\
\phi_i = -n_{\rm H} \frac{D_{z}}{Sc} \frac{\partial}{\partial z}\left(\frac{n_i}{n_{\rm H}}\right),  \label{eq:phi}
\end{align}
where $P_i$, and $L_i$ represent the production rate and the loss rate, respectively, of species $i$.
The second term in the left-hand side of Equation (\ref{eq:basic}) describes diffusion in turbulent disks. 
The vertical diffusion coefficient for gas and very small dust grains (i.e well-coupled to gas) is assumed to be
\begin{equation}
D_z = \langle \delta v^2_z\rangle/\Omega,
%D_z = \alpha_z c_s^2/\Omega,
\end{equation}
where $\Omega$ is the Keplerian orbital frequency \citep[][]{fromang06,okuzumi11}.
The vertical velocity dispersion, $\langle \delta v^2_z\rangle$, is assumed to be  
\begin{equation}
\langle \delta v^2_z\rangle =\alpha_z c_s^2(z),
\end{equation}
where $c_s$ is the local sound speed.
The Schmidt number (ratio of gas to dust diffusivity) is
\begin{equation}
Sc \sim 1+(\Omega\tau_{\rm s})^2,
\end{equation}
where $\tau_{\rm s}$ is the stopping time of dust grains \citep[][]{youdin07}.
We adopt $Sc = 1$, since the term $\Omega\tau_{\rm s}$ is much less than unity in the midplane of our disk model (e.g., $\sim$10$^{-6}$ at $r=5$ AU).
In other words, gaseous species and dust grains (grain surface species) have the same diffusivity.
It is worth noting that in our approach, only the mean composition of ice mantle of grains is obtained at each spatial point in the disk.
In reality the ice composition could vary among grains in the same spatial grid at a given time, since the motion of grain is random in turbulent gases, 
and since grains have different thermal and irradiation history \citep{ciesla12}.   
%The vertical velocity dispersion, $\langle \delta v^2_z\rangle$, is assumed to be  
%\begin{equation}
%\langle \delta v^2_z\rangle = \alpha_z c_s^2,
%\end{equation}

The activity of MRI depends on the ionization degree.
MRI can be stabilized near the midplane of disks, where the ionization degree is low and magnetic field is decoupled to plasma.
It is called dead zone \citep[e.g.,][]{gammie96,sano00}.
Three-dimensional isothermal MHD simulations in the shearing box approximation show $\langle \delta v^2_z\rangle$ in the dead zone 
is smaller than that in the MRI active layer by more than one order of magnitude, depending on the strength of the magnetic field \citep{okuzumi11,gressel12}.
In our fiducial models, however, we assume that $\alpha_z$ is constant for simplicity.
We run three models in which $\alpha_z$ is set to be 0, 10$^{-3}$, and 10$^{-2}$.
The effect of dead zone is discussed in Section \ref{sec:deadzone}.

%where $n_{\rm mid}$ and $z_{\rm crit}$ are the number density of hydrogen nuclei at the midplane and the height where $R_{M} = 100$, respectively.
%To reduce the diffusion coefficient in dead zone, we introduce the artificial factor, $\chi$, which equals $10^{-2}$ 
%at the midplane and 1 at $z=z_{\rm crit}$.
%Between the midplane and $z=z_{\rm crit}$, $\chi$ has a value linearly interpolated in the $n^{-1}$ space, 
%following \citet{okuzumi11}, in which they showed $n\langle \delta v^2_z\rangle$ is almost constant in the vertical direction at disk core.
Equation (\ref{eq:basic}) is integrated for 10$^6$ yr using implicit finite differencing on a linear grid consist of vertical 60 points at a specific radius.
Our code is based on Nautilus code \citep{hersant09}, but we improved the treatment of turbulent diffusion to 
account for the full coupling of mixing with chemistry according to \citet{heinzeller11}.
As boundary conditions, we assume that there is no flux through the midplane and the upper boundary of the disk.
We integrate Equation (\ref{eq:basic}) at 35 radial points from $r = 1$ AU to $r = 300$ AU.
We ignore radial accretion and radial mixing in the present study; we will investigate them in forthcoming papers.

%Therefore, this study would show the maximum potential of the vertical mixing to change the D/H ratio in water ice.
%In the calculation, the diffusion coefficient is evaluated at every time step, using latest $x_e$.  

%$\alpha$ would depend on $R$ and $z$..
%The absorption line of $^{12}$CO is saturated at cloud surface, while $^{13}$CO is photo dissociated in deeper regions. 

%% In this section, we use  the \subsection command to set off
%% a subsection.  \footnote is used to insert a footnote to the text.


%% This section also includes several of the displayed math environments
%% mentioned in the Author Guide.

\section{RESULT \label{sec:result}}
\subsection{Effect of Vertical Mixing on Oxygen Chemistry \label{sec:abundance}}
Beyond the snow line ($r \gtrsim 1$ AU in our models), the bulk of water exists as ice on grain surfaces.
In this subsection, we describe how the vertical mixing affects abundances of oxygen reservoirs beyond the snow line in the disk.
As an example, we look into the results at $r=10$ AU and 30 AU in the model with $\alpha_z = 10^{-2}$.
In the following, species X on grain surfaces are denoted as X$_{\rm ice}$.

\begin{figure}
\epsscale{1.0}
\plotone{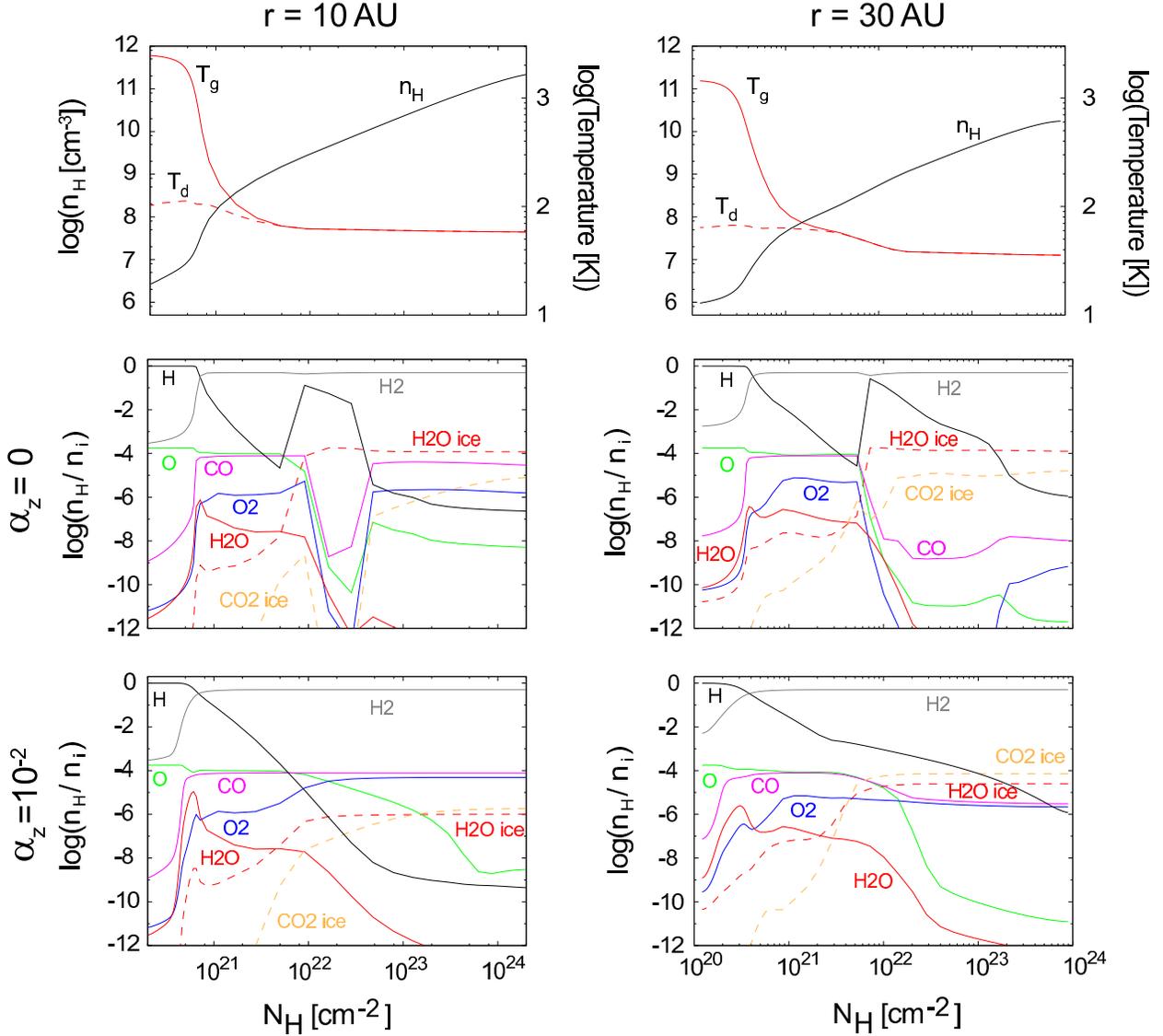}
\caption{Physical parameters (top panels), abundances of selected species in the model with $\alpha_z = 0$ (middle panels), and with 
$\alpha_z = 10^ {-2}$ (bottom panels) as functions of vertical column density at $r=10$ AU (left panels) and 30 AU (right panels) at $t=10^6$ yr.
The solid lines represent gas-phase species, while the dashed lines represent ice-mantle species.
\label{fig:vd_ab}}
\end{figure}

Figure \ref{fig:vd_ab} shows physical parameters (top panels), abundances of assorted O-bearing species in the model with $\alpha_z = 0$ (middle panels), and $\alpha_z = 10^ {-2}$ (bottom panels) as functions of vertical column density at $r=10$ AU (left panels) and 30 AU (right panels), at $t=10^6$ yr.
In the model without mixing, the disk can be divided into two layers in terms of the main oxygen reservoir; atomic oxygen in the disk surface and water ice near the midplane.
Above the O/H$_2$O$_{{\rm ice}}$ transition, photoreactions on grain surfaces are efficient, and the water ice abundances are low ($<$10$^{-6}$).
Thermal desorption of water ice is not efficient either at $r=10$ AU or $r=30$ AU, since dust temperatures are less than the sublimation temperature of water ice ($\sim$150 K) even at the disk surface.

%Photodissociation of water vapor also contributes to OH formation above the O/H$_2$O-ice transition, since OH has the smaller photodissociation cross section at the Ly-$\alpha$ wavelength than that of water vapor by a factor of nine.
%we note ion-neutra path 'Í'»'ñ'Èo'ðwater'Élock'·'é'Ù'ÇŒø—¦"I'Å'È'¢(branching)
%At $r=100$ AU, water vapor forms via the ion-neutral reactions, and shows the abundances of 10$^{-8}$--10$^{-7}$ at $N_{\rm H} \lesssim 10^{22}$ cm$^{-2}$.
%Photodesorption does not significantly contribute water vapor formation at $r=100$ AU in our model.

%The position of the transition zone of the two layers is mainly determined by the attenuation length of UV radiation.
%As shown in Table 1, most part of oxygen is initially in water in our models.
%In the atomic oxygen layer water ice is not stable and is efficiently destroyed by subsequent photodissociation on grain surfaces, followed by thermal desorption of atomic oxygen, 

\begin{figure}
\epsscale{0.8}
\plotone{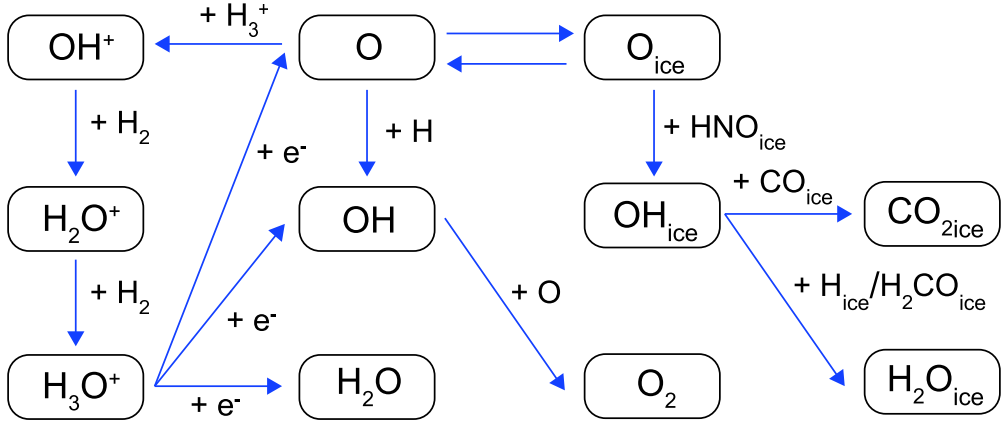}
\caption{Important reaction routes to (re)form water ice, O$_2$, and CO$_2$ ice from atomic oxygen beyond the snow line.
\label{fig:network}}
\end{figure}

\begin{figure}
\epsscale{0.6}
\rotatebox{90}{\plotone{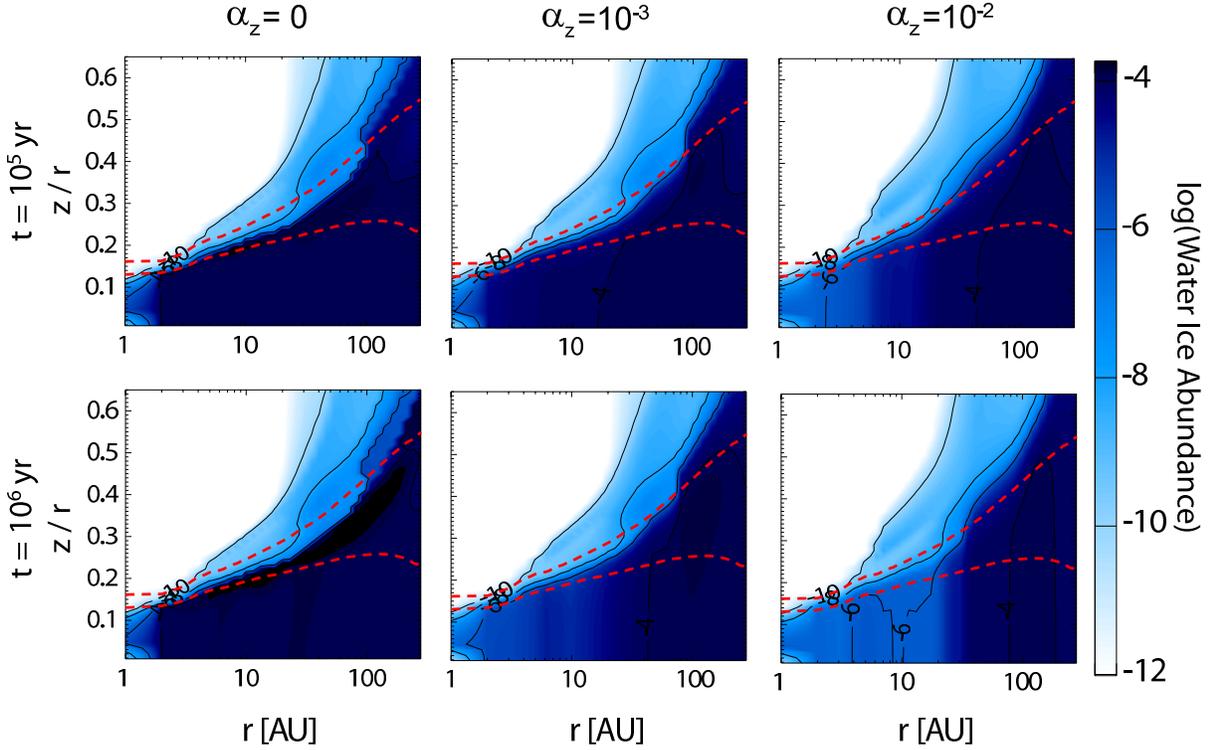}}
\caption{Spatial distributions of water ice abundance in the model with $\alpha_z$ = 0 (left panels), 10$^{-3}$ (middle panels), and 10$^{-2}$ (right panels) at 10$^5$ yr (top panels) and 10$^{6}$ yr (bottom panels). The vertical axes represent height normalized by the radius. The dashed lines indicate vertical visual extinction of unity and ten. \label{fig:2d_h2o_ice_ab}}
\end{figure}

In the model with mixing, water ice is transported to the surface through the O/H$_2$O$_{{\rm ice}}$ transition.
Such water ice is destroyed by photoreactions to produce atomic oxygen.
On the other hand, atomic oxygen is transported to the deeper layers in the disk, and is mainly converted to O$_2$ at $r=10$ AU, and CO$_2$ ice and water ice at 30 AU.
Then, the water ice abundance near the midplane decreases with time; 
it is less than the canonical value of $\sim$10$^{-4}$ at 10$^6$ yr especially at $r=10$ AU.
At $r \gtrsim 40$ AU, on the other hand, water ice abundance is $\sim$10$^{-4}$ even in the model with mixing (see Figure \ref{fig:2d_h2o_ice_ab}), 
since most of atomic oxygen is converted back to water ice.
The variation of the major O-bearing species, which are (re)formed from the atomic oxygen, partly depends on dust temperatures near the O/H$_2$O$_{{\rm ice}}$ transition 
(strictly speaking the height $z^{\ast}$; see Section \ref{sec:timescale}), where the conversion of atomic oxygen to other species mainly takes place.
In Figure \ref{fig:network} we present main formation paths to (re)form O$_2$, CO$_2$ ice, and water ice beyond the snow line in our models.
We note here that the formation reactions of these species include OH as the reactant:
\begin{align}
{\rm O} + {\rm OH} & \rightarrow {\rm O_2} + {\rm H}, \label{react:o+oh}\\
{\rm CO_{ice}} + {\rm OH_{ice}} & \rightarrow {\rm CO_{2ice}} + {\rm H_{ice}}, \label{react:co+oh}\\
{\rm H_{ice}} + {\rm OH_{ice}} & \rightarrow {\rm H_2O_{ice}} + {\rm H_{ice}}. \label{react:h+oh}\\
{\rm H_2CO_{ice}} + {\rm OH_{ice}} & \rightarrow {\rm H_2O_{ice}} + {\rm HCO_{ice}}. \label{react:h2co+oh}
\end{align}
Reaction (\ref{react:o+oh}) is a gas-phase reaction, while the others are grain-surface reactions.
In order for water ice to be efficiently reformed, OH should be mainly formed on grain surfaces, and water ice formation should be more efficient than CO$_2$ ice formation.

%In addition to reaction (\ref{react:h+oh}), water ice also forms via the following reaction channels on grain surfaces, which begin from the reaction including OH:
%\begin{align}
%{\rm O_{\,ice}} + {\rm OH_{\,ice}} & \rightarrow {\rm HO_{2\: ice}}, \label{react:o+oh(ice)}\\
%{\rm HO_{2\: ice}} + {\rm H_{\,ice}} & \rightarrow {\rm H_2O_{2\: ice}},\\
%{\rm H_2O_{2\: ice}} + {\rm H_{\,ice}} & \rightarrow {\rm H_2O_{\,ice}} + {\rm OH_{\,ice}}.\\
%\end{align}
%Although our reaction network includes other surface reactions to form water ice, they are less efficient than above reactions 
%at $r=10$ AU and 30 AU.

At $r=10$ AU, dust temperatures are too high ($\sim$60 K) to effectively form OH on grain surfaces; adsorbed atomic oxygen is likely to be evaporated 
before it meets other reactive species.
Hence, the main formation route of OH from atomic oxygen is the gas phase reactions: radiative association with atomic hydrogen and/or a sequence of ion-neutral reactions, 
followed by the recombination of H$_3$O$^+$ with an electron. 
In the recombination of H$_3$O$^+$ with an electron, the branch to produce OH accounts for 74 \% of the total recombination rate \citep{jensen00}.
Although the recombination of H$_3$O$^+$ also has a branch to produce water vapor, it accounts only for 25 \% of the total rate.
Once OH is formed in the gas phase, it mostly reacts with atomic oxygen and is converted to O$_2$ before being adsorbed onto dust grains, 
due to the high abundance of atomic oxygen ($\gtrsim$10$^{-5}$) around the O/H$_2$O$_{{\rm ice}}$ transition.
The ratio of the O$_2$ formation rate and adsorption rate of OH is expressed as
\begin{equation}
\begin{split}
\frac{k_1n_{\rm O}n_{\rm OH}}{\pi a^2 v_{\rm th}n_{\rm d}n_{\rm OH}} = & 400\left(\frac{x_{\rm O}}{10^{-5}}\right)\left(\frac{k_1}{4\times10^{-11}\,\,{\rm cm^3\,\,s^{-1}}}\right) \\
         & \times \left(\frac{a}{0.1\,\,\mu{\rm m}}\right)^{-2}\left(\frac{v_{\rm th}}{10^4\,\,{\rm cm\,\,s^{-1}}}\right)^{-1}\left(\frac{x_d}{10^{-12}}\right)^{-1},
\end{split}
\end{equation}
where $k_1$, the rate coefficient of OH + O, is $4\times10^{-11}$ cm$^3$ s$^{-1}$ independent of the gas temperature.
Then, the conversion to O$_2$ is dominant.
Although the value of $k_1$ has some uncertainties (2--8$\times10^{-11}$ cm$^3$ s$^{-1}$) according to KIDA 
\citep[http://kida.obs.u-bordeaux1.fr; see also][]{wakelam12,hincelin11}, the ratio exceeds unity even if we use the lowest value.
We note that conversion of O$_2$ to water on grain surfaces (O$_{\rm 2ice}$ $\xrightarrow[]{{\rm H_{ice}}}$ HO$_{\rm 2ice}$ $\xrightarrow[]{{\rm H_{ice}}}$ 
H$_2$O$_{\rm 2ice}$ $\xrightarrow[]{{\rm H_{ice}}}$ H$_2$O$_{\rm ice}$) \citep{miyauchi08}
is not efficient due to high dust temperature.
Another reason for the high abundance of O$_2$ is that its destruction rate by UV photons is small compared to that of 
water ice, since Ly$\alpha$ cross section of O$_2$ is lower than that of water ice by about two orders of magnitude \citep{van dishoeck06,mason06}.
%The O$_2$ abundance is as high as 10$^{-6}$--10$^{-5}$ even above the O/H$_2$O$_{{\rm ice}}$ transition in the model without mixing.
%It leads to the flatter distribution of O$_2$ compared to that of water ice in the model with mixing.
%Then, the net upward flux of O$_2$ from the midplane to the surface, where photodissociation is efficient, is smaller than that of water ice (see Equation \ref{eq:phi}). 

At $r\gtrsim30$ AU, lower dust temperatures ($\lesssim$40 K) allow OH to mainly form on grain surfaces.
Water ice is mainly formed by H$_{\rm ice}$ + OH$_{\rm ice}$ at $r\lesssim 30$--40 AU.
CO and atomic hydrogen are mostly in the gas phase at these radii.
Since CO has a higher desorption energy than atomic hydrogen (Table \ref{table:edes}), 
adsorbed CO stays for a longer time on grain surfaces and% ($\propto \exp(E_{\rm des}/T_{\rm d})$).
%Although the hopping rate ($\propto \exp(-0.5E_{\rm des}/T_{\rm d})$) of CO is lower than that of atomic hydrogen and the H/CO abundance ratio in the gas phase is less than unity, 
%the longer desorption timescale compensates for them; 
CO$_{\rm ice}$ + OH$_{\rm ice}$ proceeds faster than H$_{\rm ice}$ + OH$_{\rm ice}$.
CO$_2$ ice has a smaller photoabsorption cross section than that of water ice, which further ensures the high abundance of CO$_2$ ice at $r \sim 30$ AU.
H$_2$CO has a higher desorption energy than CO, and mostly exists as ice at $r\gtrsim30$ AU.
At $r \gtrsim 40$ AU, H$_2$CO$_{\rm ice}$ + OH$_{\rm ice}$ is more efficient than H$_{\rm ice}$ + OH$_{\rm ice}$ and CO$_{\rm ice}$ + OH$_{\rm ice}$, due to high abundance of H$_2$CO ice near the O/H$_2$O$_{\rm ice}$ transition ($\gtrsim$10$^{-8}$).
%We will discuss the uncertainties of grain surface chemistry in Section \ref{sec:uncertainty}.

\begin{figure}
\epsscale{0.85}
\plotone{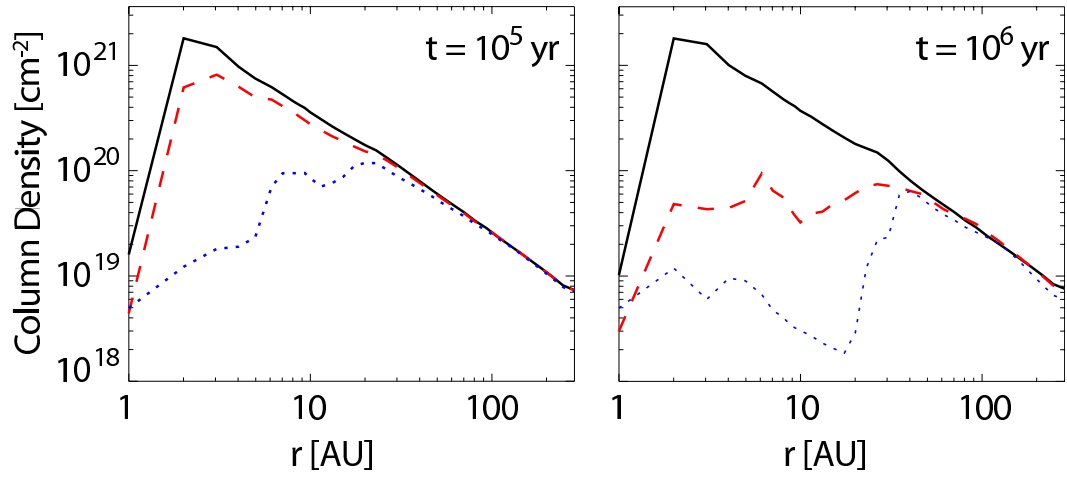}
\caption{Radial distributions of water ice column density at 10$^5$ yr (left panel) and 10$^6$ yr (right panel) in the model with $\alpha_z=0$ (solid lines), 10$^{-3}$ (dashed lines), and 10$^{-2}$ (dotted lines). \label{fig:column_ice}}
\end{figure}

Figure \ref{fig:2d_h2o_ice_ab} shows two-dimensional distributions of the abundance of water ice, 
while Figure \ref{fig:column_ice} shows radial distributions of its column density.
In the model without mixing, the water ice abundance beyond the snow line stays constant.
In the model with mixing, the water ice abundance near the midplane decreases.
The tendency is more significant at smaller radius, since reformation of water ice is prohibited by high dust temperatures, and since the timescale of mixing is small 
(see Section \ref{sec:timescale} for quantitative discussions).
Above the O/H$_2$O$_{{\rm ice}}$ transition, a fraction of oxygen exists as water vapor, and its abundance reaches a maximum near the H/H$_2$ transition.
In the regions with the gas temperature $T \gtrsim 300$ K, water vapor is formed via the neutral-neutral reaction
\begin{align}
%{\rm O} + {\rm H_2} & \rightarrow {\rm OH} + {\rm H}, \label{react:o+h2} \\ 
{\rm OH} + {\rm H_2} & \rightarrow {\rm H_2O} + {\rm H}, \label{react:oh+h2}
\end{align}
which has the potential energy barrier of 1740 K.
Water vapor is also formed by a sequence of ion-neutral reactions, followed by the recombination of H$_3$O$^+$ with an electron, 
which is the dominant formation path in the regions with gas temperature $T \lesssim 300$ K.

In the model without mixing, we can see the local bumps of the abundances of atomic hydrogen at $N_{\rm H} \sim 10^{22}$ cm$^{-2}$.
The positions correspond to where dust grains start to be covered with ice and chemisorption sites become unavailable (i.e. formation rate of hydrogen molecule on grain surfaces decreases).
These bumps are smoothed out via mixing.
In the model with mixing, peak abundance of water vapor is enhanced in the disk atmosphere, since the H$_2$ abundance above the H/H$_2$ transition 
increases \citep[Reaction \ref{react:oh+h2}; see also][]{heinzeller11}.

%\begin{equation}
%\tau_{{\rm mix}, i} = \frac{\int^{z_{\rm tr}}_{0}n_{i}dz'}{\phi_i(z_{\rm tr})} \sim 10^5 \;[{\rm yr}]
%\end{equation}
\subsection{Destruction Timescale of Water Ice \label{sec:timescale}}
\begin{figure}
\epsscale{0.7}
\plotone{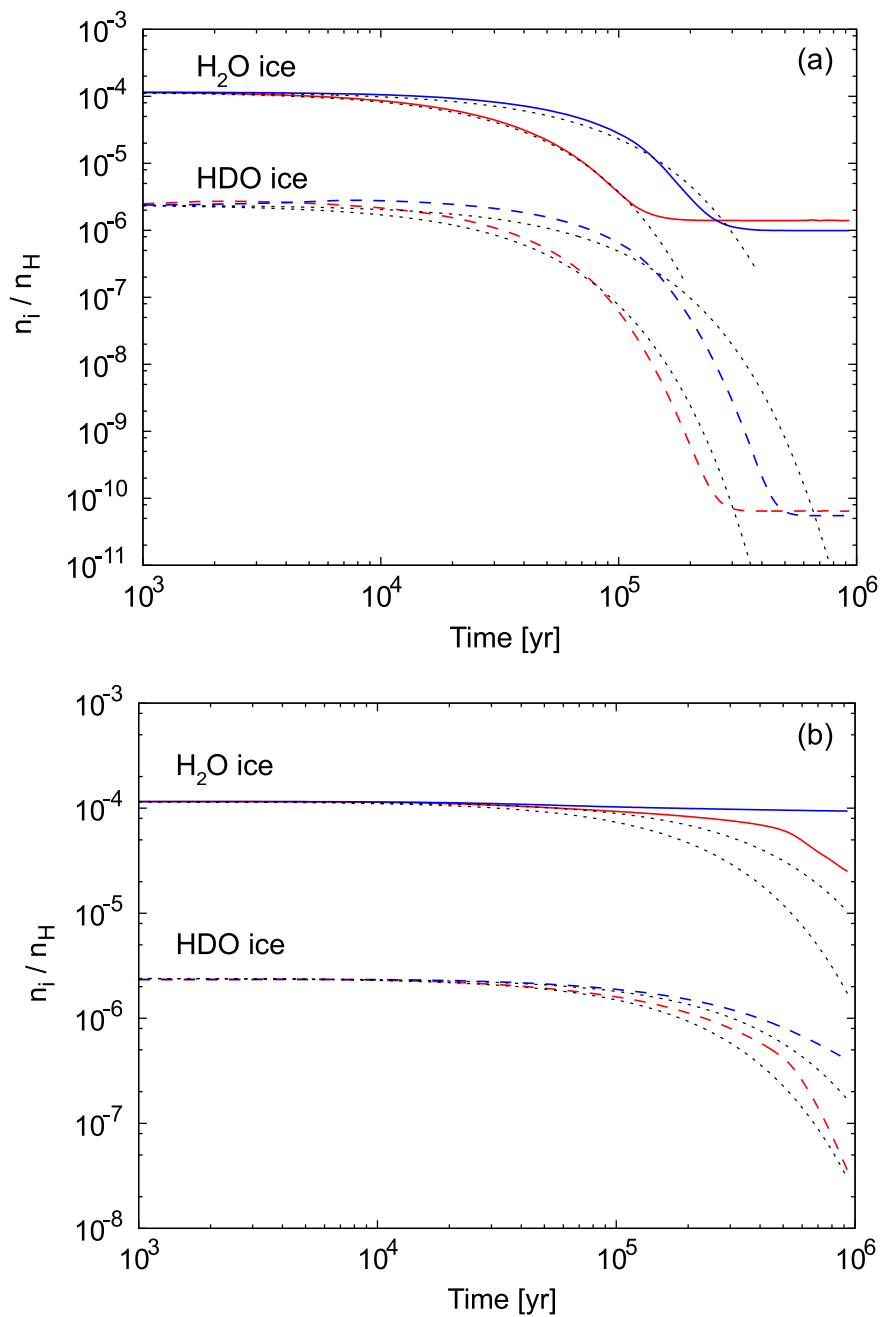}
\caption{Temporal variations of abundances of H$_2$O ice (solid lines) and HDO ice (dashed lines) in the model with $\alpha_z = 10^{-2}$ at the midplane of (a) $r=5$ AU (red) and 10 AU (blue), and (b) $r=30$ AU (red) and 50 AU (blue). The dotted lines depict the temporal variation given by Equation (\ref{eq:xice}). \label{fig:xice}}
\end{figure}

When turbulence exists, water ice near the midplane is destroyed via the combination of vertical mixing and photoreactions.
This process significantly changes oxygen chemistry as mentioned above.
The destruction timescale of water ice would be determined by the timescale of vertical transport via mixing,
which in turn, can be estimated from the column density of water ice divided by the upward flux $\phi_{\rm H_2O_{\rm ice}}(z^{\ast})$:
\begin{equation}
\tau^{\rm mix}_{\rm H_2O_{\rm ice}} = \frac{N_{\rm H_2O_{\rm ice}}}{\phi_{\rm H_2O_{\rm ice}}(z^{\ast})}. \label{eq:t_mix} %= \frac{N_{\rm H_2O\,ice}}{n(z_{\rm tr})D_{z}(z_{\rm tr})}\left(\frac{\partial x_{\rm H_2O\,ice}}{\partial z}\right)^{-1}_{\rm z_{\rm tr}}
\end{equation}
We define $z^{\ast}$ as the height where $\partial \phi/\partial z$ reaches a minimum (i.e. where the net upward flux of water ice reaches a maximum).
We note that the precise evaluation of the timescale requires the solution of Equation (\ref{eq:basic}), since the flux $\phi$ is proportional to the composition gradient.
We evaluate $\tau^{\rm mix}_{\rm H_2O_{\rm ice}}$ using our numerical data at $t = 10^4$ yr in the model with mixing.
The upper panel of Figure \ref{fig:xice} shows temporal variation of water ice abundance in the model with $\alpha_z = 10^{-2}$ at the midplane of $r=5$ AU and 10 AU.
It shows that the water ice abundance at the midplane at $t<$ a few 10$^5$ yr in the numerical simulations are well reproduced by the exponential decay:
\begin{equation}
x_{\rm H_2O_{\rm ice}}(t) = x_{\rm H_2O_{\rm ice}}(t=0)\exp \left(-\frac{t}{\tau^{\rm mix}_{\rm H_2O_{\rm ice}}}\right). \label{eq:xice}
\end{equation}
It ensures that $\tau^{\rm mix}_{\rm H_2O_{\rm ice}}$ represents the characteristic timescale of destruction of water ice.
At $t >$ a few 10$^5$ yr, the water ice abundance is significantly low compared to the initial value, 
and destruction and formation almost balance.
The lower panel of Figure \ref{fig:xice} is similar to the upper panel, but shows the results at $r=30$ AU and 50 AU.
At these radii, Equation (\ref{eq:xice}) is no more valid, since water ice reformation is not negligible,
although Equation (\ref{eq:t_mix}) should be still valid.

We can estimate the destruction timescale of HDO ice from Equation (\ref{eq:t_mix}), but with the subscript replaced by HDO ice.
We find that the timescales of HDO ice and H$_2$O ice are similar, because their chemistry is similar.
For example, $\tau^{\rm mix}_{\rm H_2O_{\rm ice}}$ and $\tau^{\rm mix}_{\rm HDO_{\rm ice}}$ are $6.3\times10^4$ yr and $5.7\times10^4$ yr, respectively, at $r=10$ AU.
Then we simply denote them as $\tau^{\rm mix}$ in the rest of the paper.

\begin{figure}
\epsscale{0.85}
\plotone{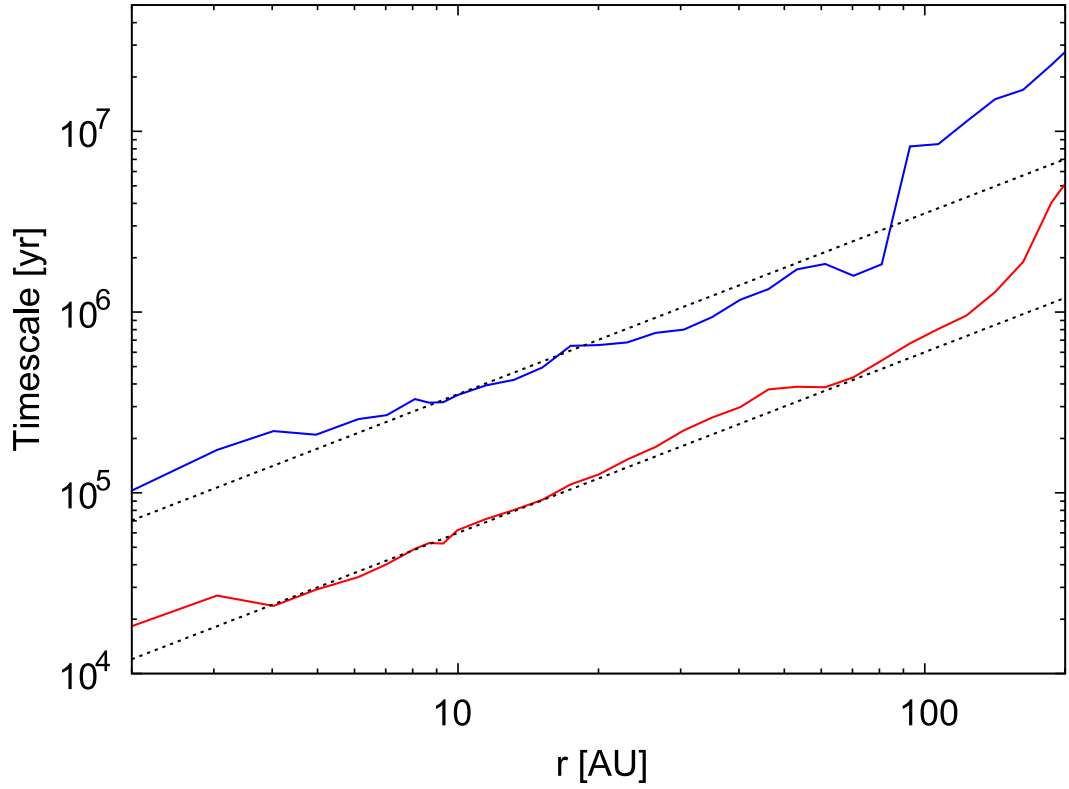}
\caption{Radial variations of the destruction timescale of water ice, $\tau^{\rm mix}$, in the model with $\alpha_z=10^{-3}$ (blue solid line) and 10$^{-2}$ (red solid line). The dotted lines indicate Equation (\ref{eq:t_mix_r}). \label{fig:t_mix}}
\end{figure}

In Figure \ref{fig:t_mix}, we present radial variations of $\tau^{\rm mix}$.
We fit the numerical data at 2 AU $\lesssim r \lesssim$ 100 AU and find that $\tau^{\rm mix}$ depends almost linearly on the radial distance from the central star:
\begin{equation}
\tau^{\rm mix}(r) \sim 6\times10^4\left(\frac{r}{10\,{\rm AU}}\right)\left(\frac{\alpha_z}{10^{-2}}\right)^{-0.8} \,\,{\rm yr} \,\,\,\,(2\,{\rm AU} \lesssim r \lesssim 100\,{\rm AU}). \label{eq:t_mix_r}
\end{equation}
%This linear radial dependence probably comes from the radial dependence of $N_{\rm H}$ ($\propto r^{-1}$) and the photon fluxes ($\propto r^{-2}$).
%The dependence of $\tau^{\rm mix}$ on the photon flux through $\phi$ may be natural, since ices transported to the surface are destroyed by photoreactions.
The timescale is smaller in the inner regions.
At $r \gtrsim 100$ AU, destruction timescale is not determined by $\tau^{\rm mix}$, since photoreactions rather than the upward transport of water ice limits the destruction rate, 
which is clear from the flat distribution of water ice abundance in the vertical direction (Figure \ref{fig:2d_h2o_ice_ab}).

\subsection{Deuteration of Water Ice \label{sec:waterice}}
We describe how the vertical mixing affects deuterium chemistry, focusing on the largest deuterium reservoir, HD.
In our initial abundance for the disk model, which is set by the collapsing core model, $\sim$30 \% of deuterium is in species other than HD 
and D$_2$ (Table \ref{table:initial}).
For example, $\sim$15 \% of deuterium is in water, and its D/H ratio is $\sim \!\! 2 \times 10^{-2}$.
In the collapsing core, the species which are mostly formed in a cold era ($\sim$10 K), such as water, are enriched in deuterium through the isotope exchange reactions, 
such as, H$_3^+$ + HD $\rightleftharpoons$ H$_2$D$^+$ + H$_2$ + 230 K.
The backward reaction is endothermic, and is negligible compared to the forward reaction in the low gas temperature.
Then, H$_3^+$ and several other species which are subject to the direct exchange reactions become enriched in deuterium.
Chemical reaction network propagates the enrichment to other species, so that a significant amount of deuterium is delivered to the species 
other than HD and D$_2$ \citep[e.g.,][]{aikawa12}.

\begin{figure}
\epsscale{0.8}
\plotone{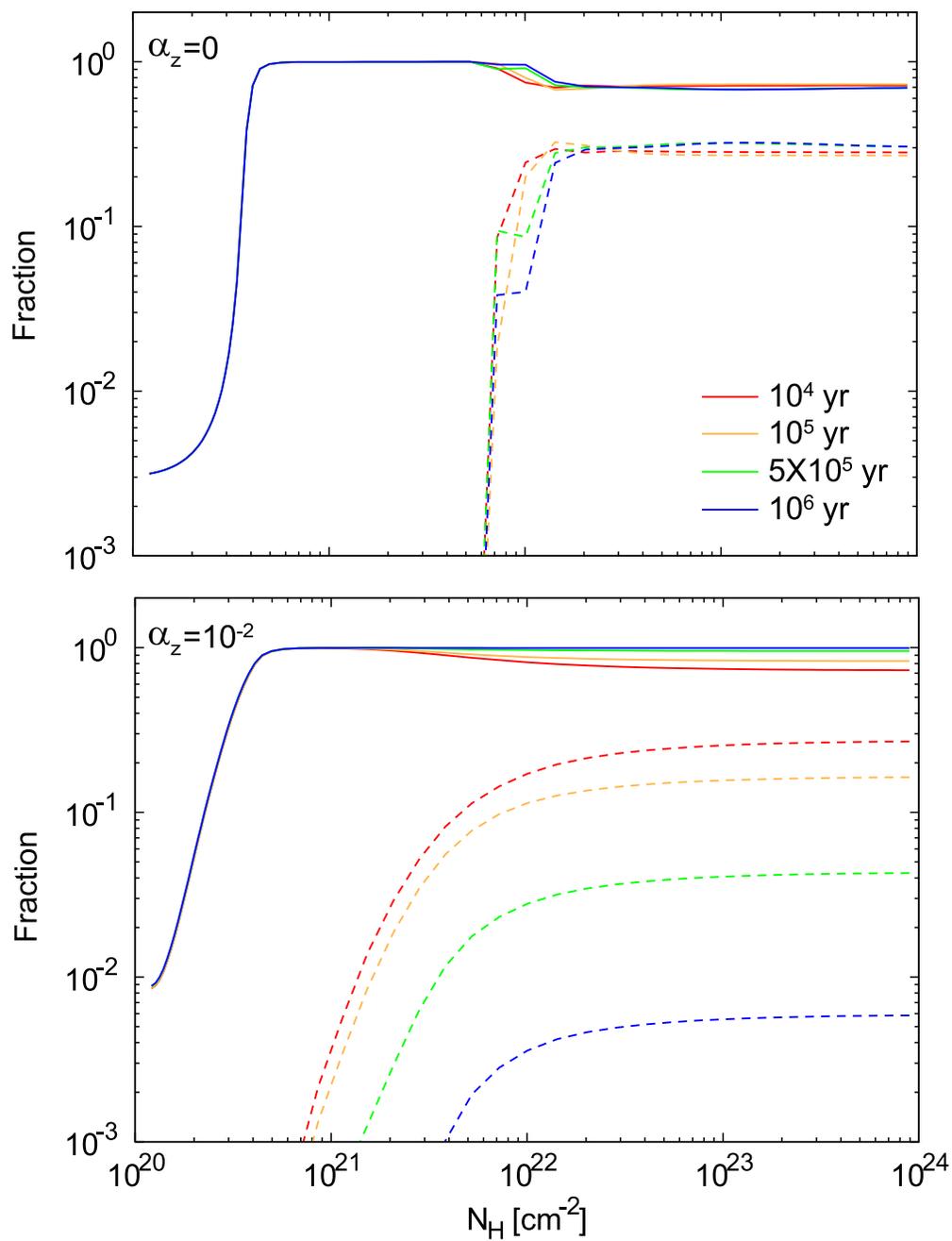}
\caption{Temporal variations of the vertical profile of a fraction of elemental deuterium in the form of HD and D$_2$ (solid lines) and the total fraction in other molecular form than HD and D$_2$ (dashed lines) at $r=30$ AU in the model with $\alpha_z=0$ (upper panel) and $\alpha_z=10^{-2}$ (lower panel). \label{fig:xhd}}
\end{figure}

Figure \ref{fig:xhd} shows vertical profiles of a fraction of elemental deuterium in the form of HD and D$_2$ (i.e. $(x_{\rm HD} + 2x_{\rm D_2})/1.5\times10^{-5}$) 
in the disk models at $r=30$ AU.
The total fraction in other molecular form than HD and D$_2$, mainly ice, are also shown in Figure \ref{fig:xhd}.
At $N_{\rm H} \sim 10^{20}$ cm$^{-2}$, these fractions are small, since deuterium is mostly in atomic form.
In the model without mixing, almost all deuterium is in HD (and D$_2$) at $N_{\rm H} = 10^{21}$--10$^{22}$ cm$^{-2}$, since the other major deuterium 
reservoirs, such as HDO ice, are mostly destroyed by photoreactions.
In the deeper regions, on the other hand, those reservoirs survive for 10$^6$ yr, since the timescales of photoreactions are 
longer than 10$^{6}$ yr there (see Figure \ref{fig:photorate}).
In the model with mixing, the deuterated ices are transported to the surface to be destroyed by photoreactions, while HD is transported to the midplane.
Note that the total elemental abundances in gas and dust, are not changed by the mixing, as long as dust grains are dynamically well-coupled to the gas.
Because of the higher gas temperatures in the disk than that in the cold core, the endothermic exchange reactions, such as H$_2$D$^+$ + H$_2$, are not negligible.
Therefore, reformation of deuterated ices is not efficient enough to compensate for the destruction.
As a result, the HD abundance near the midplane increases with time, while the abundances of the other deuterium reservoirs decrease.
For example, $\sim$20 \% of deuterium is in the species other than HD at $t =10^5$ yr in the midplane of $r=30$ AU, 
while the fraction decreases to less than 1 \% at $t=10^6$ yr.
 
%The reason for the decrease of the HDO$_{{\rm ice}}$/H$_2$O$_{{\rm ice}}$ ratio comes from higher gas temperatures of the disk than that of the cold core.
%In other words, the D/H ratio of reforming water ice is less than $10^{-2}$.

As shown in the bottom panel of Figure \ref{fig:xice}, the H$_2$O ice abundances are almost constant in the midplane at $r=30$ AU and 50 AU, 
since reformation of H$_2$O ice (partly) compensates for the destruction.
On the other hand, the HDO ice abundances decrease in a similar timescale to $\tau^{\rm mix}$.
Then, the HDO$_{{\rm ice}}$/H$_2$O$_{{\rm ice}}$ ratio decreases in a similar timescale to $\tau^{\rm mix}$ at these radii.
%It is the consequence of the HDO$_{{\rm ice}}$/H$_2$O$_{{\rm ice}}$ reformation rate ratio is $<$10$^{-2}$.
In the inner radii, $r=5$ AU and 10 AU, reformation of both ices is negligible, due to high dust temperatures.
Since the destruction timescales of H$_2$O and HDO ices are similar, substantial changes of the HDO$_{{\rm ice}}$/H$_2$O$_{{\rm ice}}$ ratio does not occur 
until their abundances decrease significantly and reach the steady state ($t>$ a few 10$^5$ yr; upper panel of Figure \ref{fig:xice}).
Therefore, $\tau^{\rm mix}$ gives the approximate timescale of the decline of the HDO$_{\rm ice}$/H$_2$O$_{\rm ice}$ ratio at $r \gtrsim 30$ AU in our models, 
while it gives the timescale of H$_2$O ice decrease at $r \lesssim 30$ AU.

\begin{figure}
\epsscale{0.85}
\plotone{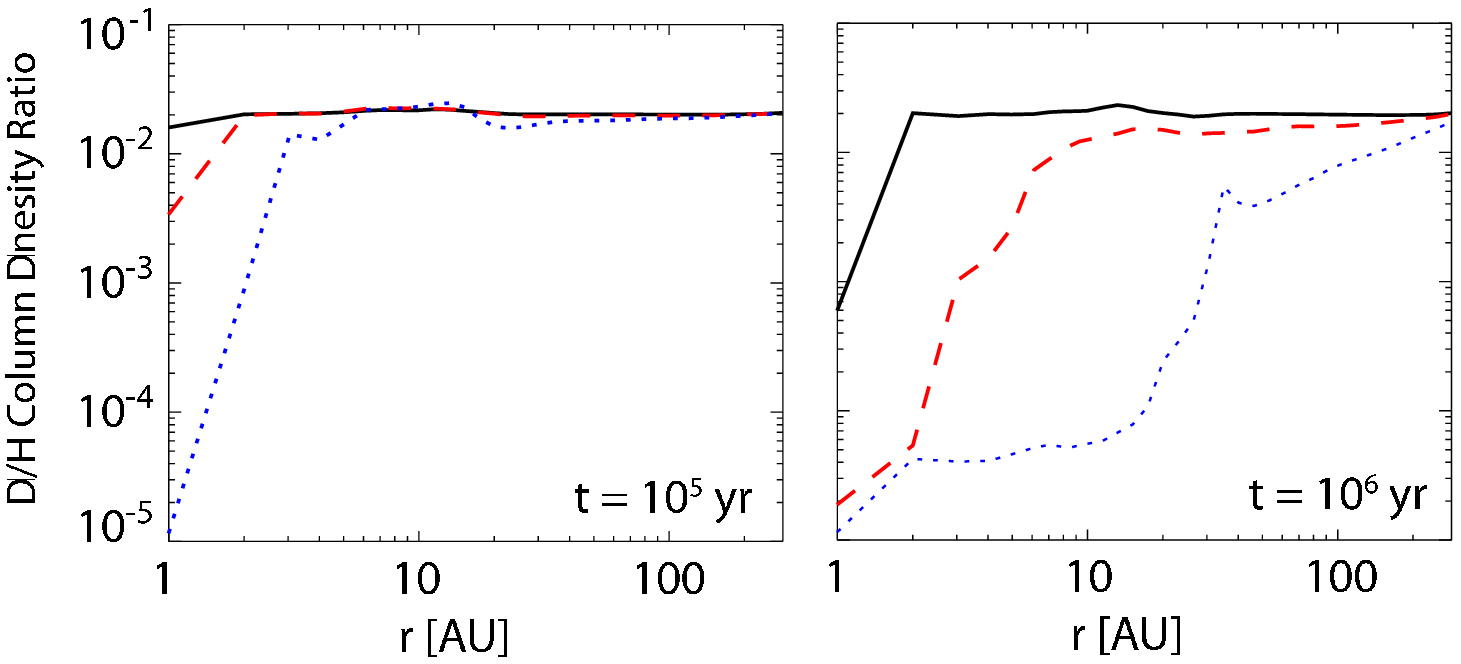}
\caption{Radial distributions of HDO$_{{\rm ice}}$/H$_2$O$_{{\rm ice}}$ column density ratio at 10$^5$ yr (left panel) and 10$^6$ yr (right panel) in the model with $\alpha_z=0$ (solid lines), 10$^{-3}$ (dashed lines), and 10$^{-2}$ (dotted lines). \label{fig:column_dh}}
\end{figure}

Figure \ref{fig:column_dh} shows radial variations of the HDO$_{\rm ice}$/H$_2$O$_{\rm ice}$ column density ratio.
While the D/H ratio retains its initial value ($\sim \!\! 2 \times 10^{-2}$) for 10$^6$ yr in the model without mixing, 
it decreases with time in the model with mixing.
Since $\tau^{\rm mix}$ is smaller and the gas temperature is higher in inner radii, the resultant D/H ratio of water ice column density is 
roughly an increasing function of radius.
In the model with $\alpha_z = 10^{-3}$, the HDO$_{{\rm ice}}$/H$_2$O$_{{\rm ice}}$ ratio decreases by more than one order of magnitude at $r \lesssim 3$ AU within 10$^6$ yr.
In the model with more efficient mixing, $\alpha_z=10^{-2}$, such a significant decline of the HDO$_{{\rm ice}}$/H$_2$O$_{{\rm ice}}$ ratio can be seen 
inside $r=2$ AU and $30$ AU, within 10$^5$ yr and 10$^6$ yr, respectively.
It should be noted, however, that in these regions with the low HDO$_{{\rm ice}}$/H$_2$O$_{{\rm ice}}$ ratio, water ice abundance and its column density are also lower 
than those in the model without mixing by about two orders of magnitude, and water ice is no longer the major oxygen reservoir.

In summary, at $r<30$ AU, warm gas temperatures lead to low HDO$_{{\rm ice}}$/H$_2$O$_{{\rm ice}}$ ratio, 
while warm dust temperatures hamper water ice reformation.
At $r \gtrsim 30$ AU, the HDO$_{{\rm ice}}$/H$_2$O$_{{\rm ice}}$ ratio decreases by up to one order of magnitude within 10$^6$ yr,
without significant decrease of the water ice column density.

%Now, we know how much the D/H ratio in water ice could decrease via vertical mixing within 10$^6$ yr at a maximum, i.e., when the abundance of H$_2$O ice remains constant, while that of HDO ice decrease in a timescale of $\tau^{\rm mix}$.
%Under the condition, the D/H ratio in water at the time, $t$, would be expressed as the initial ratio times $\exp(-t/\tau^{\rm mix})$.
%Decreasing the ratio by more than one order of magnitude could occur at $r<80$ AU (20 AU), while decreasing by more than two orders of magnitude could occur at $r<30$ AU (3 AU) in the model with $\alpha_z=10^{-2}$ (10$^{-3}$).

%Since $z^{\ast}$ is mainly determined by the balance of photolysis and upward flux of water ice, $z^{\ast}$ (i.e. $\tau_{\rm mix}$) depends on the strength of vertical mixing ($\alpha_z$) and the intensity of FUV photons.

\subsection{Deuteration of Water Vapor \label{sec:vapor}}
\begin{figure}
\epsscale{0.6}
\rotatebox{90}{\plotone{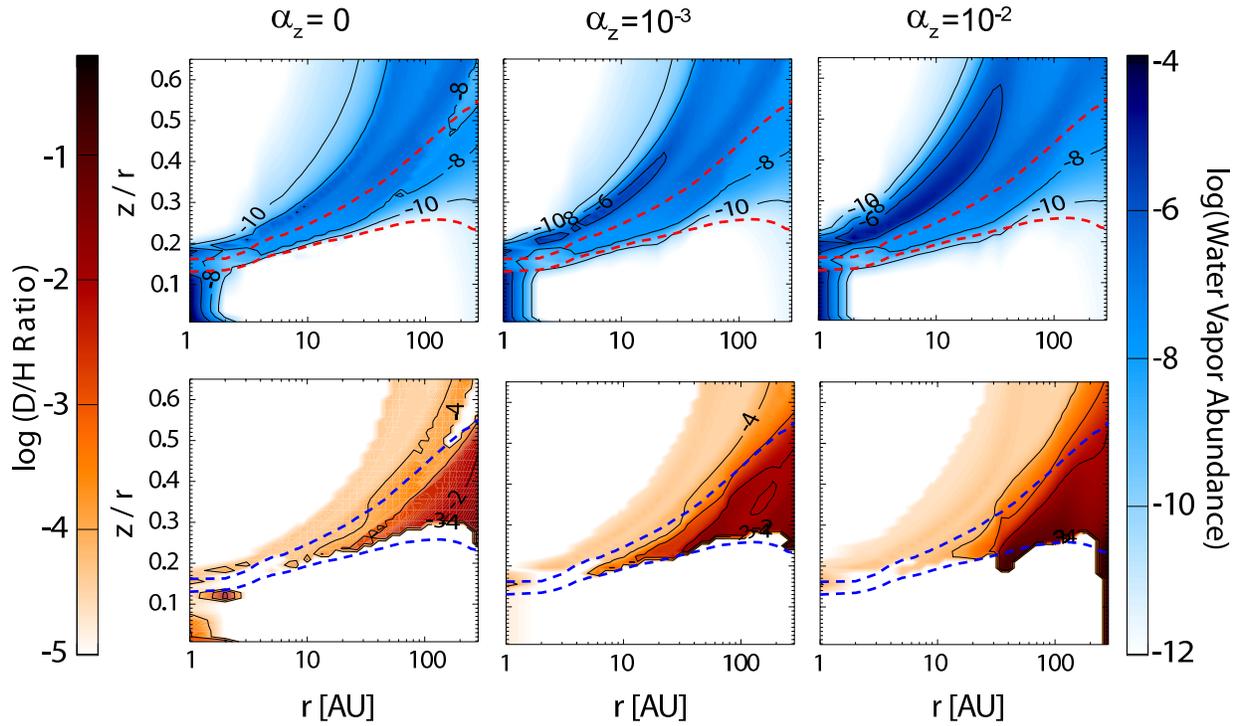}}
\caption{Spatial distributions of water vapor abundance (blue colors) and its D/H ratio (orange colors) in the model with $\alpha_z$ = 0 (left panels), 10$^{-3}$ (middle panels), and 10$^{-2}$ (right panels) at 10$^{6}$ yr. The dashed lines indicate vertical visual extinction of unity and ten. The D/H ratios are only shown in regions where the water vapor abundance is higher than 10$^{-12}$.\label{fig:2d_h2o_gas}}
\end{figure}

Figure \ref{fig:2d_h2o_gas} shows spatial distributions of water vapor abundance and the gaseous HDO/H$_2$O ratio.
In our model, water vapor is moderately abundant ($\gtrsim$10$^{-8}$) in three regions: (i) the midplane inside the snowline, 
(ii) the disk surface at $r < 50$ AU, and (iii) the outer disk.
They have been identified by previous disk chemical models without mixing \citep{woitke09b,meijerink12b}.
This structure is conserved in the model with vertical mixing.
In this subsection, we discuss how the D/H ratio of water vapor is determined in these regions.

%{\it The midplane inside the snowline} {\it (}$T_{\rm d} \sim 200$ K, $A_{\rm V} > 10$, {\it and} $r \lesssim 1$ AU{\it )}.
{\it (i) The midplane inside the snowline} {\it  ($T_g \sim T_d \sim$ 200 K, $A_V >$ 10, and $r \sim$ 1 AU)}.

At $r \sim 1$ AU, the bulk of water exists in the gas phase near the midplane, 
since dust temperatures are higher than the sublimation temperature of water.
Since FUV photons and X-rays are strongly attenuated, H$_2$O and HDO are mainly destroyed by reactions with ions and photons induced by cosmic-rays in a timescale of a few 10$^5$ yr.
We note that cosmic-ray is also slightly attenuated; while the unattenuated cosmic ray ionization rate is $5\times10^{-17}$ s$^{-1}$, 
the ionization rate is $\sim \!\! 1\times10^{-17}$ s$^{-1}$ in the midplane of $r=1$ AU.
H$_2$O and HDO reform via proton (deuteron) transfer (e.g., H$_3$O$^+$ + NH$_3$) and/or neutral-neutral reactions (OH + H$_2$ and OD + H$_2$; see Table \ref{table:react}).
This cycle reduces the HDO/H$_2$O ratio, since the products of HDO destruction, such as H$_2$DO$^+$, are converted not only to HDO, but also to H$_2$O.
The HDO/H$_2$O ratio decreases from the initial value in a similar timescale to the destruction timescale, 
and reaches $\sim$10$^{-3}$ at 10$^6$ yr in the model without mixing.
In the model with mixing, their destruction timescales are shortened by a combination of the upward transport and photoreactions at the disk surface.
The HDO/H$_2$O ratio decreases in a timescale of $1 \times 10^4$ yr ($7 \times 10^4$ yr) in the model with $\alpha_z = 10^{-2}$ (10$^{-3}$), and finally reaches $\sim$10$^{-5}$.

{\it (ii) The inner disk surface} {\it ($T_g \gtrsim$ 300 K, $A_V <$ 1, and $r \lesssim$ 50 AU)}.

The HDO/H$_2$O ratio is as low as or less than 10$^{-4}$ in this region, where dust and gas temperatures are decoupled.
This result agrees with \citet{willacy09}.
The HDO/H$_2$O ratio is similar to the OD/OH ratio, since H$_2$O and HDO is mainly formed via OH + H$_2$ and OD + H$_2$, respectively.
While the exchange reaction of OH + D deuterates OH, the reverse reaction with the activation barrier of 810 K, OD + H, 
prevents the significant deuteration of OH (i.e. water) at $T_{\rm g} \gtrsim 300$ K.
Vertical mixing increases the abundance of water vapor in this region by enhancing the H$_2$ abundance (Section \ref{sec:abundance}), 
while it does not strongly affect the HDO/H$_2$O ratio.

{\it (iii) The outer disk} {\it ($T_g \lesssim$ 300 K, $A_V <$ a few, and $r \gtrsim$ 50 AU)}.

In the region with $T_{\rm g} < 300$ K, H$_2$O and HDO mainly forms via a sequence of ion-neutral reactions, 
followed by the recombination of H$_3$O$^+$ and its isotopologues with an electron, 
while they are mainly destroyed via photodissociation and reactions with ions.
It should be noted that the abundance of water in this region depends on the X-ray flux \citep{meijerink12b}, 
and that photodesorption of water ice is important to keep a fraction of oxygen in the gas phase \citep[e.g.,][]{dominik05}.
Vertical mixing increases both the abundances of OH and atomic deuterium by a factor of a few.
The abundance of OD also increases, since it is mainly formed by the reaction of OH + D.
Since H$_2$DO$^+$ is formed via a sequence of ion-neutral reactions, initiated by the reactions between OD and ions, the HDO/H$_2$O ratio in this region also increases in the model with mixing.

\begin{figure}
\epsscale{0.6}
\plotone{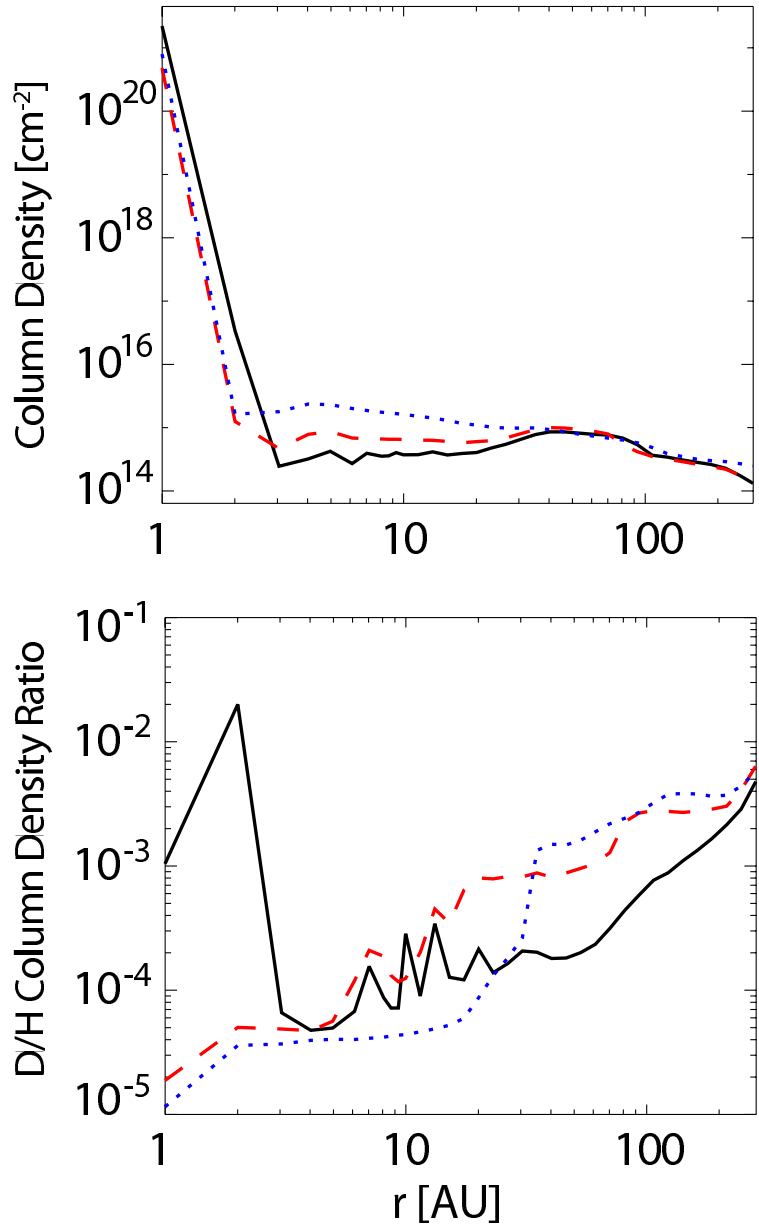}
\caption{Radial variations of water vapor column density (top panel) and HDO/H$_2$O column density ratio (bottom panel) at 10$^6$ yr in the model with $\alpha_z=0$ (solid lines), 10$^{-3}$ (dashed lines), and 10$^{-2}$ (dotted lines). \label{fig:column_gas}}
\end{figure}

In Figure \ref{fig:column_gas}, we present radial variations of water vapor column density and its D/H ratio.
At 3 AU $\lesssim r \lesssim$ 30 AU, the column densities of water vapor increase by up to one order of magnitude via vertical mixing compared to the model without mixing.
This is due to the increased abundance of hot water ($T_g \gtrsim$ 300 K).
The column density at $\sim$1 AU decreases via vertical mixing by a factor of a few, since a fraction of oxygen transported from the surface is converted to O$_2$.
At the inner disks ($< 1$ AU), where gas temperature near the midplane is higher than 300 K, reformation of water vapor via Reaction (\ref{react:oh+h2}) would be faster than O$_2$ formation.

Vertical mixing also affects the HDO/H$_2$O column density ratio.
One of the most important effects of the mixing is a substantial decrease of the ratio inside the snow line.
Beyond the snow line, on the other hand, vertical mixing decreases or increases the column density ratio, 
depending on which dominates in the column density, hot water ($T_g \gtrsim$ 300 K) or warm water ($T_g \lesssim$ 300 K).
The ratio at $r=2$ AU is high ($\sim$10$^{-2}$) in the model without mixing, since a fraction of water exists 
in the gas phase ($\sim$10$^{-6}$) near the midplane via thermal desorption of ice.

Finally, we point out that the D/H ratios of water vapor and ice are different, especially in the inner regions ($r \lesssim 30$ AU), regardless of the strength of vertical mixing.
This suggests that it is difficult to constrain the D/H ratios of water ice in the midplane of the inner disks from the observations 
of D/H ratios of water vapor in protoplanetary disks.

\subsection{Comparisons to Other Works}
The effect of mixing on water chemistry beyond the snow line ($r > 10$ AU) was studied by \citet{semenov11}, considering both radial and vertical mixing.
Their chemical network is also based on \citet{garrod06}.
They found that water ice column densities in the model with and without mixing differ only by a factor of $<$2--5, which is consistent with our results at $r \gtrsim 30$ AU.
However, in the inner disks, column densities of water ice decreases by more than one order of magnitude in our models.
The difference seems to mainly come from the higher dust temperatures in our models compared to those in \citet{semenov11}.
Radial mixing may also help maintain the high abundance of water ice, which will be investigated in forthcoming papers.

While there are several studies focusing on deuterium fractionation in PPDs, none of them has considered mixing.
Here, we briefly compare our model without mixing to those of \citet{willacy09}, which focused on the inner disks ($r \lesssim 30$ AU) considering the radial accretion.
They found that both beyond and inside the snow line, the D/H ratio of water in the midplane retains its initial value ($\sim$10$^{-2}$) for 10$^6$ yr.
In our models without mixing, the D/H ratio of water ice in the midplane retains the initial value beyond the snow line, while the D/H ratio of water vapor decreases 
inside the snow line in a timescale of a few 10$^5$ yr.
The different results inside the snow line seem to come from higher unattenuated cosmic-ray ionization rate in our models ($5\times10^{-17}$ s$^{-1}$),
than that in \citet{willacy09} ($1.3\times10^{-17}$ s$^{-1}$).
The timescale of destruction and reformation of water vapor, and thus the timescale of decline of its D/H ratio is inversely proportional to the cosmic-ray ionization rate.
If we adopt the unattenuated cosmic-ray ionization rate of $1.3\times10^{-17}$ s$^{-1}$, the timescale becomes comparable to or longer than 10$^6$ yr.
The combination of radial accretion and evaporation of water ice at the snow line may also help maintain the initial D/H ratio of water.
\citet{willacy09} also found that the HDO/H$_2$O ratio in the inner disk surface is as low as $10^{-4}$, which is consistent with our model. 

\section{DISCUSSION}
\subsection{Initial HDO/H$_2$O Ratio \label{sec:initial_ratio}}
Some hot corino sources show the HDO/H$_2$O ratio as high as the initial value in our fiducial models \citep[$\sim$10$^{-2}$; ][]{taquet13a}, 
while some sources show much lower value \citep[$10^{-4}$--$10^{-3}$; ][]{jorgensen10,persson13}.
Here, we discuss the dependence of our results presented in Section \ref{sec:result} on the initial HDO/H$_2$O ratio.

In order to make the molecular composition with the low HDO/H$_2$O ratio, we artificially decrease the abundances of deuterated species 
except for HD by a factor of forty compared to the initial abundance used in our fiducial models.
We reset the HD abundance so that the elemental deuterium abundance of $1.5\times10^{-5}$ relative to hydrogen.
For species without deuterium, we use the same abundances as those in our fiducial models.
Then, the elemental abundances used here are slightly different from those in our fiducial models (at most $\sim$4 \%), except for deuterium.
In this initial abundance, the HDO/H$_2$O ratio is $\sim \!\! 5 \times 10^{-4}$, which is similar to the cometary value ($\sim \!\! 6 \times 10^{-4}$), 
and $\sim$99 \% of deuterium is in HD.

\begin{figure}
\epsscale{0.55}
\plotone{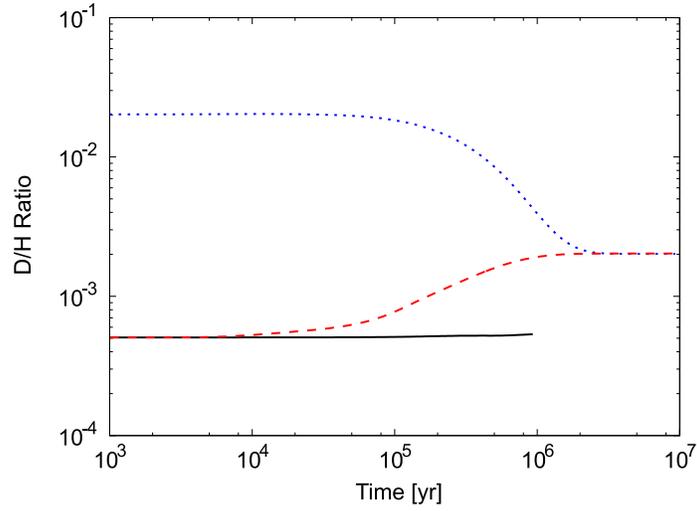}
\caption{Temporal variations of HDO$_{{\rm ice}}$/H$_2$O$_{{\rm ice}}$ ratio in the midplane of $r=50$ AU in the model with $\alpha_z=10^{-2}$ and the intial ratio of $\sim \!\! 2 \times 10^{-2}$ (dotted line) and $\sim \!\! 5 \times 10^{-4}$ (dashed line).
The solid line indicates the HDO$_{{\rm ice}}$/H$_2$O$_{{\rm ice}}$ ratio in the model without mixing and with the initial ratio of $\sim \!\! 5 \times 10^{-4}$. \label{fig:d_h_4}}
\end{figure}

In Figure \ref{fig:d_h_4}, we show temporal variations of HDO$_{{\rm ice}}$/H$_2$O$_{{\rm ice}}$ ratio in the midplane of $r=50$ AU.
The ratio increases with time from $5\times10^{-4}$ to $2\times10^{-3}$ in 10$^6$ yr in the model with $\alpha_z=10^{-2}$, 
while the ratio is nearly constant in the model without mixing.
Unlike the fiducial model, transport of deuterium via mixing is not important, since almost all deuterium is in HD from the beginning.
Still, transport of oxygen affects water and deuterated water chemistry;
atomic oxygen is transported from the surface to the deeper region, and (re)form H$_2$O and HDO ices.
The HDO/H$_2$O ratio increases, since the D/H ratio of reformed water ice is larger than $5\times10^{-4}$.
After a few 10$^6$ yr, the HDO$_{{\rm ice}}$/H$_2$O$_{{\rm ice}}$ ratio reaches the steady-state value of $2\times10^{-3}$, which is independent of the initial ratio.
The steady-state value should correspond to the D/H ratio of reformed water ice in the disk at a given radius (i.e. temperature).

\begin{figure}
\epsscale{0.55}
\plotone{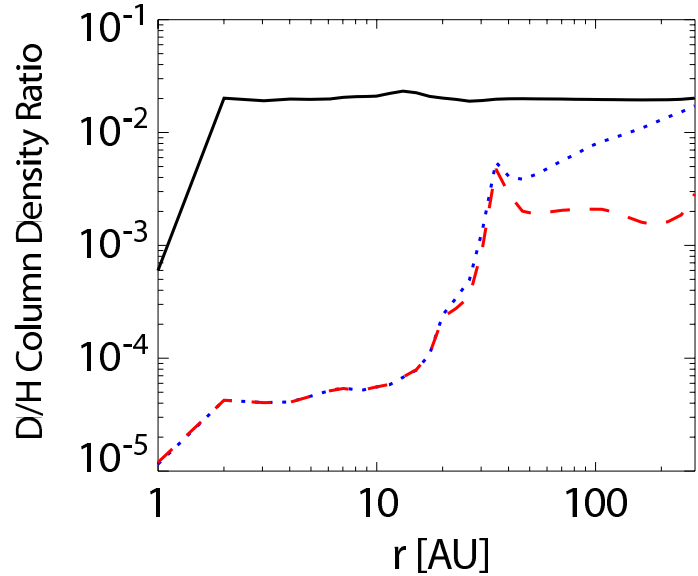}
\caption{Radial distributions of HDO$_{{\rm ice}}$/H$_2$O$_{{\rm ice}}$ column density ratio at $t=10^6$ yr in the model with $\alpha_z=10^{-2}$ and the initial ratio of $\sim \!\! 2 \times 10^{-2}$ (dotted line) and $\sim \!\! 5 \times 10^{-4}$ (dashed line). The parameter $\alpha_z$ is set to be 10$^{-2}$ in both models. The solid line indicates the HDO$_{{\rm ice}}$/H$_2$O$_{{\rm ice}}$ ratio in the model without mixing and with the initial ratio of $\sim \!\! 2 \times 10^{-2}$.\label{fig:d_h_4_r}}
\end{figure}

Figure \ref{fig:d_h_4_r} shows radial distributions of HDO$_{{\rm ice}}$/H$_2$O$_{{\rm ice}}$ column density ratio at $t=10^6$ yr.
In the inner region ($r \lesssim 40$ AU), where $\tau^{\rm mix}$ is much smaller than 10$^6$ yr, the HDO$_{{\rm ice}}$/H$_2$O$_{{\rm ice}}$ ratio is independent of the initial ratio.
In the outer region, on the other hand, the ratio depends on the initial ratio at least for $10^6$ yr.

%Since destruction timescales of H$_2$O and HDO ices are similar, if the HDO$_{{\rm ice}}$/H$_2$O$_{{\rm ice}}$ reformation rate ratio is $<$10$^{-2}$, 
%the HDO$_{{\rm ice}}$/H$_2$O$_{{\rm ice}}$ abundance ratio decreases with time.
%The opposite is also true (see Section \ref{sec:initial_ratio}).
%In our fiducial models, the HDO$_{{\rm ice}}$/H$_2$O$_{{\rm ice}}$ abundance ratio decreases with time, as shown in Figure \ref{fig:2d_h2o_ice_dh}.

%we integrate rate equations for 10$^{7}$ yr under $T_{\rm g} = T_{\rm d} = 30$ K,
%$n_{\rm H} = 10^4$ cm$^{-3}$, and $A_{\rm V} = 10$ instead of the collapsing core model.
%We choice dust temperature of 30 K to avoid the freezing-out of CO onto grain surfaces, while we choice $t=10^7$ yr to obtain 
%the substantial H$_2$O ice abundance ($\sim10^{-4}$).
%CO is the main destroyer of H$_2$D$^+$, and it prohibits the significant deuteration of gas and ice.

\subsection{Uncertainties of Chemistry on Grain Surfaces \label{sec:uncertainty}}
\subsubsection{Water Formation}
Recently, \citet{meijerink12b} investigated water formation on carbonaceous grain surfaces, considering chemisorption of atomic hydrogen.
They found that the reaction between physisorbed atomic oxygen and chemisorbed atomic hydrogen is very efficient at $T_{\rm d}<40$ K, 
but the efficiency declines rapidly at higher temperatures.
It also should be noted that chemisorption sites are unavailable on grain surfaces with ice-mantles (i.e. when $x_{\rm H_2O_{\rm ice}} > N_{\rm site}x_{\rm d} \sim 10^{-6}$).
Therefore, the inclusion of water ice formation on chemisorption sites would not significantly change our results.

\begin{figure}
\epsscale{0.6}
\plotone{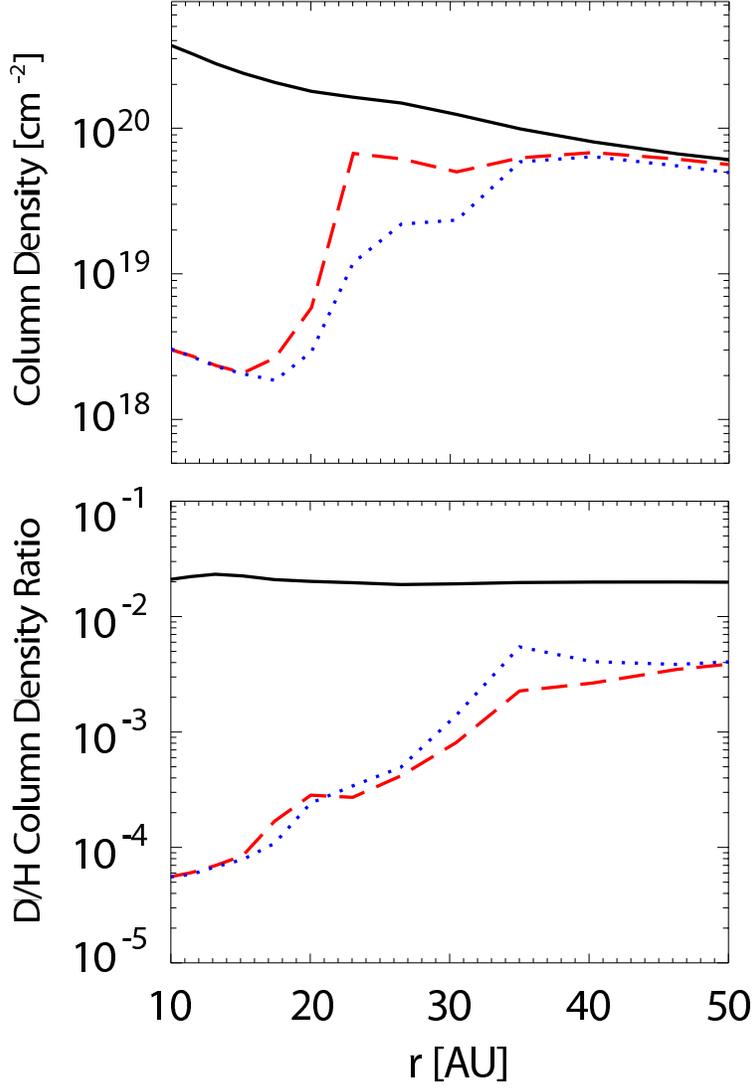}
\caption{Radial distributions of water ice column density (upper panel) and HDO$_{{\rm ice}}$/H$_2$O$_{{\rm ice}}$ column density ratio at $t=10^6$ yr in the model with $\alpha_z=10^{-2}$. The dashed lines indicate the model with $E_{\rm des} =600$ K for atomic hydrogen, while the dotted lines indicate the model with $E_{\rm des} =450$ K (our fiducial model). The parameter $\alpha_z$ is set to be 10$^{-2}$ in both models. The solid lines indicate the values in the model without mixing. \label{fig:r600}}
\end{figure}

In our fiducial models, the desorption energy of atomic hydrogen is fixed to be 450 K, following \citet{garrod06}.
\citet{watanabe10} and \citet{hama12} observed various kinds of potential sites (or site distributions) with different depths ($E_{\rm des}$) for atomic 
hydrogen on amorphous water ice in their experiments.
According to molecular dynamics simulations, the site distribution peaks at $E_{\rm des} \sim 400$ K and $\sim$600 K for crystalline 
and amorphous water ices, respectively \citep{al-halabi07}.
Higher desorption energy of atomic hydrogen would lead to the longer residence time on grain surfaces, and thus a higher (re)formation rate of water ice.
In order to check the dependence of $E_{\rm des}$ on our results, we reran our model from $r = 10$ AU to $r = 50$ AU with $E_{\rm des} = 600$ K 
for atomic hydrogen and deuterium.
The resultant radial distribution of water ice column density and its D/H ratio are shown in Figure \ref{fig:r600}.
In the model with $E_{\rm des} = 600$ K, the water ice reformation compensates the destruction at $r \gtrsim 20$ AU, 
while in our fiducial model ($E_{\rm des} = 450$ K), it compensates the destruction only at $r \gtrsim 40$ AU.
In the inner region ($r \lesssim 20$ AU), the significant reduction of water ice column density occurs, even with $E_{\rm des} = 600$ K.
The radial variations of the HDO$_{\rm ice}$/H$_2$O$_{\rm ice}$ column density ratio does not strongly depends on $E_{\rm des}$.
If the $E_{\rm des}$ is 600 K, the column density ratio of HDO$_{\rm ice}$/H$_2$O$_{\rm ice}$ reaches the cometary value in 10$^6$ yr, 
while the H$_2$O ice abundance is kept high, in the comet forming regions ($r \sim 20$--30 AU)(see Section \ref{sec:cometary_water}).

In order to efficiently (re)form water ice, OH should be formed on grain surfaces as discussed in Section \ref{sec:abundance}.
In our model, OH$_{\rm ice}$ is mainly formed via the reaction,
\begin{equation}
{\rm O_{\rm ice}} + {\rm HNO_{\rm ice}} \rightarrow {\rm OH_{\rm ice}} + {\rm NO_{\rm ice}}.
\end{equation}
The simplest reaction, O$_{\rm ice}$ + H$_{\rm ice}$, is less efficient, because of the lower desorption energy (or shorter residence time on grain surfaces) of atomic hydrogen than that of HNO (Table \ref{table:edes}). 
The product NO$_{\rm ice}$ cycles back to HNO$_{\rm ice}$ via the reaction,
\begin{equation}
{\rm NO_{\rm ice}} + {\rm H_{\rm ice}} \rightarrow {\rm HNO_{\rm ice}}, \label{react:no+h} 
\end{equation}
which is faster than O$_{\rm ice}$ + H$_{\rm ice}$, because of the longer residence time of NO than that of atomic oxygen.
There is a caveat about this loop of NO$_{\rm ice}$/HNO$_{\rm ice}$ interconversion as pointed out by \citet{garrod09}.
The grain-surface network of \citet{garrod06}, which we employed, has only a limited number of reactions involving NO$_{\rm ice}$ and HNO$_{\rm ice}$.
While the network would be appropriate for cold environments (10 K), where atomic hydrogen is the dominant mobile reactant, 
NO$_{\rm ice}$ and HNO$_{\rm ice}$ could be subject to various other reactions in warmer environments, where species with heavy elements also become mobile.
\citet{garrod09} mentioned that a more comprehensive reaction network that allowed NO$_{\rm ice}$ to react with, for example, atomic oxygen
and nitrogen (which the current network does not) would make such loop inefficient.
%In order to check the importance of the NO$_{\rm ice}$/HNO$_{\rm ice}$ interconversion for water ice formation at warm temperatures, 
%we reran the models without Reaction (\ref{react:no+h}).
%We confirmed that the O$_2$ dominated region is beyond $r =30$ AU
However, it is unclear whether the caveat applies to the disk chemistry with mixing.
The physical condition of the disk, such as density, is significantly different from those of clouds/cores considered by \citet{garrod09}. 
Mixing enhances the abundances of reactive species, such as atomic hydrogen and oxygen.
Modeling of surface reaction on warm dust grains with a more comprehensive reaction network, which is currently unavailable, is desirable. 

\subsubsection{Photodissociation rates of ices}
In our fiducial models, we calculate the photodissociation rates of ices (Equation \ref{eq:rphs}), assuming that only the upper most one monolayer can be dissociated,  
while photoproducts immediately recombine in the deeper layers.
We may underestimate the photodissociation rates of ices, if the probability of recombination of photofragments in the deeper layers, 
which would depend on ice temperatures \citep{oberg09c}, is much less than unity.
To investigate the dependence of our results on the parameter $N_{\rm p}$, we rerun the calculation in which $N_{\rm p}$ is set to be equal to the number of monolayers ($N_{\rm layer}$).

The increased photodissociation rate of water ice enhances the conversion of water ice to CO$_2$ ice, 
since OH ice produced by water ice photodissociation are partly converted to CO$_2$ ice.
Then, the region where CO$_2$ ice becomes the dominant oxygen reservoir is 
extended to $r\sim 50$ AU in the model with $\alpha_z = 10^{-2}$, while the boundary was $r\sim 40$ AU in the fiducial model.
In our model, CO$_2$ ice is mainly formed by the reaction of CO$_{\rm ice}$ + OH$_{\rm ice}$. 
The size of the CO$_2$ ice dominated region would also depend on the value of the activation energy barrier of this reaction.
The barrier is assumed to be 80 K \citep{ruffle01} in our models, while \citet{noble11} concluded that the reaction is likely 
to have a higher barrier based on their experiments.
If we adopt a higher value, the CO$_2$ ice dominated region should become smaller.

In our fiducial model with $N_{\rm p} = 2$, we artificially switch off the D$_{\rm ice}$ + OH$_{\rm ice}$ branch of HDO ice photodissociation (see Section \ref{sec:photochem2}).
When we assume $N_{\rm p} = N_{\rm layer}$, there is no reason to switch off the branch; 
here, we assume that the branches to produce OH ice and OD ice are equally weighted.
We found that the HDO$_{\rm ice}$/H$_2$O$_{\rm ice}$ ratios at $t=10^6$ yr in the model with $\alpha_z = 10^{-2}$ are smaller than those in our fiducial model by up to 
a factor of five in the regions where water ice is the dominant oxygen reservoir ($r\gtrsim50$ AU).

\subsection{Effect of Accretion and Grain Evolution \label{sec:accretion}}
\begin{figure}
\epsscale{0.6}
\plotone{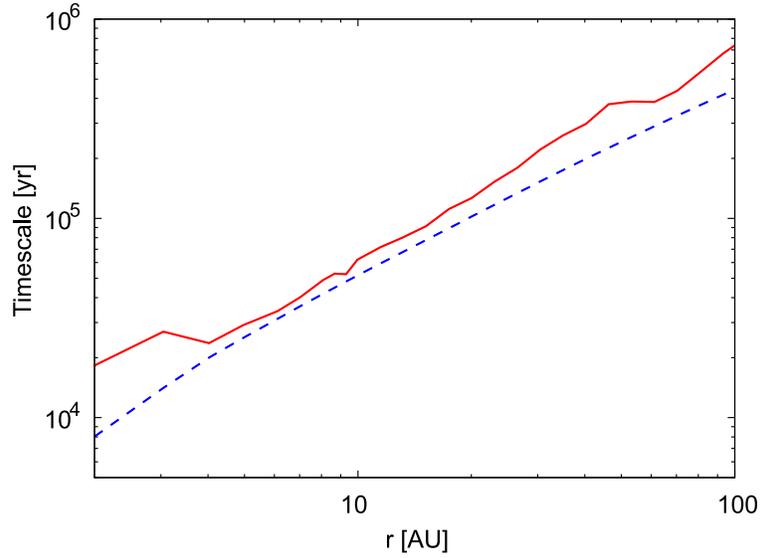}
\caption{Comparisons of mixing timescale ($\tau^{\rm mix}$; solid line) and radial accretion timescale of gas (dashed line). \label{fig:mix_vs_acc}}
\end{figure}

In the present study, we consider only vertical mixing as the mass transport in PPDs.
However, radial mass transport (in the combination with the vertical mixing) could be also important for disk chemistry
 \citep[e.g.,][] {tscharnuter07,nomura09,heinzeller11,semenov11}.
%The effect of accretion is discussed below, while the effect of radial mixing will be discussed in Section \ref{sec:cometary_water}.
In Figure \ref{fig:mix_vs_acc}, we compare $\tau^{\rm mix}$ with the timescale of radial accretion in the model with $\alpha_z = 10^{-2}$.
The accretion timescale is estimated by
\begin{alignat}{2}
\tau_{\rm acc} &= r/v_{\rm acc}, \\
v_{\rm acc} & = \dot M/2\pi r\Sigma,
\end{alignat}
where $\Sigma$ is the surface density of gases.
Note that when we determine the radial distribution of surface density in our disk models, we assume a viscous parameter of $\alpha = 10^{-2}$ as mentioned 
in Section \ref{sec:disk_structure}.
Figure \ref{fig:mix_vs_acc} shows that the accretion timescale is comparable to $\tau^{\rm mix}$.
The radial accretion would increase the water abundance and the HDO/H$_2$O ratio in our fiducial model with mixing at $r \lesssim 30$ AU, 
since they are higher at the outer radii.

We also neglect grain evolution in this work.
In our chemical models, it is assumed that 0.1 $\mu$m dust grains are uniformly distributed in the disk with the dust-to-gas mass ratio of $10^{-2}$.
In reality, PPDs contain grains as large as 1 mm, and a fraction of them would be settled to the midplane \citep[e.g.,][]{przygodda03,furlan06,furlan11}.
While the 0.1 $\mu$m grains are well-coupled to the gas, the coupling is less efficient for larger grains.
If large grains with ice mantles settle the midplane, but small grains remain in the disk surface to shield UV radiation, 
destruction of water ice via the combination of vertical mixing and photoreactions would become less efficient, 
although water vapor is subject to the destruction as long as turbulence exists.
A similar situation is considered to account for the weak emission line of water vapor detected towards TW Hya \citep{hartogh11}.

\subsection{Effect of Dead Zone} \label{sec:deadzone}
In our fiducial models, we assume that $\alpha_z$ is constant in space.
According to MHD simulations, however, velocity dispersion in a dead zone is smaller than that in a MRI active region by more than one order of magnitude, depending
on the strength of the magnetic field \citep{okuzumi11,gressel12}. 
In this subsection, we consider the models, in which $\alpha_z$ is dropped by one order of magnitude in a dead zone.
We define a dead zone as the regions with the magnetic Reynolds number, $R_{\rm M}$, is less than 100 \citep[e.g.,][]{fleming00}.
Following \citet{perez-becker11}, the magnetic Reynolds number is evaluated by 
\begin{alignat}{2}
R_{\rm M} &= c_sh/\eta, \\
\eta &= 234\sqrt{T_{\rm g}}/x'_e,
\end{alignat}
where $h$, $\eta$, and $x'_{\rm e}$ are the pressure scale height, electrical resistivity of the gas, and the electron abundance with respect to neutrals, respectively.
For example, the threshold values of $x'_{\rm e}$ for the dead zone in our disk model are $9\times10^{-13}$ and $4\times10^{-13}$ at $r=5$ AU and 10 AU, respectively.

\begin{figure}
\epsscale{0.8}
\plotone{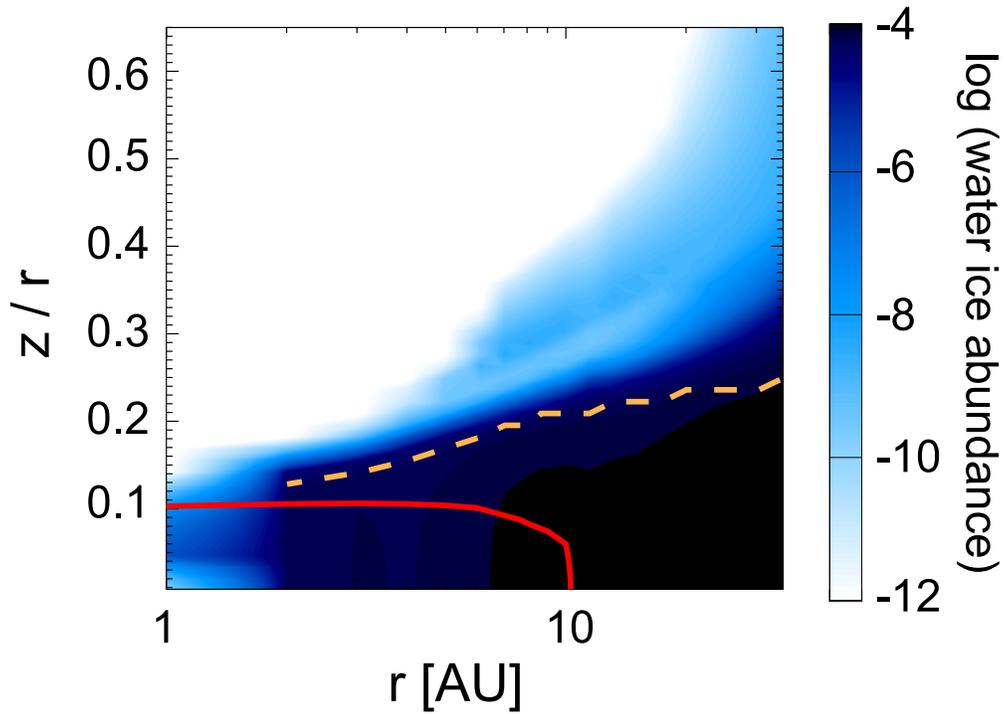}
\caption{The dead zone boundary (solid line) plotted over the distribution of water ice abundance in the model with $\alpha_z=10^{-2}$ at $t=10^4$ yr. The dashed line indicates the height $z^{\ast}$. \label{fig:deadzone}}
\end{figure}

In Figure \ref{fig:deadzone}, we show a dead zone boundary (solid line), plotted over the distribution of water ice abundance in the model with $\alpha_z=10^{-2}$ at $t=10^4$ yr.
The dead zone exists near the midplane ($z/r \lesssim 0.1$) at $r \lesssim 10$ AU.
The dashed line in Figure \ref{fig:deadzone} indicates the height $z^{\ast}$, at which the effective flux of water ice to the disk surface is determined (Section \ref{sec:timescale}).
The height $z^{\ast}$ is located above the dead zone boundary.
In other words, the water ice layer is thicker than the dead zone, and the net upward transport of water ice takes place in the MRI active region in our models.
We performed a calculation considering the dead zone, and confirmed that the existence of the dead zone does not significantly change our results on both water ice and vapor.

\subsection{Cometary Water \label{sec:cometary_water}}
Comets observed today are believed to be supplied from two distinct reservoirs, the Oort cloud and the Kuiper belt.
Until recently, it was widely accepted that the Oort cloud comets (OCCs) formed in and was scattered outward from the giant planet forming region ($r \sim 5$--30 AU),
while Jupiter-family comets (JFCs) formed in the Kuiper belt region ($r \gtrsim 30$ AU) \citep{brownlee03}.
However, recent dynamical models of solar system evolution suggest more complicated scenarios; substantial migration of the giant planets occurred and 
it led to large scale mixing of distributions of planetesimals \citep{gomes05,walsh11}.
If this is the case, OCCs and JFCs would share their origins at least in part.

So far, the HDO/H$_2$O abundance ratio has been measured in seven OCCs \citep[e.g.,][]{bockelee-morvan12} and two JFCs \citep{hartogh11,lis13}.
These observations indicate that 
\begin{enumerate}
\setlength{\parskip}{0cm}
\setlength{\itemsep}{0cm}
{\item the HDO/H$_2$O ratio of cometary water is the order of 10$^{-4}$,} 
{\item the HDO/H$_2$O ratio observed in the JFCs ((3--$<\!\!4) \times10^{-4}$) is smaller than the average value observed in the OCCs ($\sim \!\! 6 \times10^{-4}$)}.
\end{enumerate}
In our fiducial models with $\alpha_z= 0$ and $10^{-3}$, in which the initial HDO/H$_2$O ratio is $\sim \!\! 2 \times 10^{-2}$,
the model D/H ratios in water ice remain much higher than the cometary value even at $t=10^6$ yr.
In the model with $\alpha_z =10^{-2}$ and $E_{\rm des} = 600$ K for atomic hydrogen, water ice is abundant ($\sim$10$^{-4}$) and its D/H ratio is comparable to 
the cometary value at $r \sim 20$--30 AU as shown in Figure \ref{fig:r600}.
Such region with abundant water ice and the preferable D/H ratio can extend to inner radius, if the reformation of water ice is more 
efficient than that in our models.
Then, our model suggests the possibility that the D/H ratio of cometary water could be established (i.e. cometary water could be formed) in the solar nebula,
even if the D/H ratio of water ice formed in the parent molecular cloud/core was very high ($\sim$10$^{-2}$) as observed in NGC 1333-IRAS2A and NGC 1333-IRAS4A \citep{taquet13a}.
NGC 1333-IRAS4A and IRAS 16293-2422, on the other hand, show similar HDO/H$_2$O ratio to that of cometary water \citep{jorgensen10,persson13}.
Hence, it is not clear at this moment whether cometary water originates in the parent molecular cloud or the solar nebula.
Our model predicts that if the HDO/H$_2$O ratio is significantly changed in the disk, D/H ratios of other ices, such as CH$_3$OH, 
in comets would also be different from those in the molecular cloud/core.

Variation of HDO/H$_2$O ratio is another interesting issue.
If cometary water originates in molecular clouds, their HDO/H$_2$O ratio could be more uniform.
Our model predicts that the HDO/H$_2$O ratio increases with radius in the region where the model HDO/H$_2$O ratio is comparable to the cometary value.
This contradicts with the classical dynamical model in which the OCCs originated in the inner region than the JFCs, 
and may support the recent dynamical model with migration of giant planets.
It should be noted, however, that the radial gradient of HDO/H$_2$O ratio could be reversed, if grain settling proceeds faster in inner radii.
In addition, the radial distribution of the HDO/H$_2$O ratio can also be affected by the radial transport of low D/H water near the central star, 
which is not yet considered in our model \citep{yang13}.
Further studies are needed to constrain the formation regions of JHCs and OCCs from their HDO/H$_2$O ratios.

\section{CONCLUSION}
We have investigated water and deuterated water chemistry in protoplanetary disks irradiated by UV and X-ray from a central T Tauri star.
We have solved chemical rate equations with the diffusion term, mimicking the turbulent mixing in the vertical direction.
Oxygen is mainly in atomic form in the disk atmosphere, while it is in water near the midplane.
When turbulence exists, water near the midplane is transported to the disk surface and destroyed by photoreactions, while atomic oxygen is transported to
the midplane and reforms water and/or other molecules.
We found that this cycle significantly affects water and deuterated water chemistry.
Our conclusions are as follows.

\begin{enumerate}
{\item Beyond the snow line, the cycle decreases the column densities of water ice by more than one order of magnitude within 10$^6$ yr in the 
inner disk ($r\lesssim30$ AU), where dust temperatures are too high to form OH radical on grain surfaces.
Once OH radical is formed in the gas phase, it is converted to O$_2$ via the reaction of OH + O before 
it is adsorbed onto dust grains to form water ice, because of the high abundance of atomic oxygen near the O/H$_2$O$_{\rm ice}$ transition.
The outer edge of such regions moves to $r = 20$ AU, if the desorption energy of atomic hydrogen is as high as 600 K.
Our model indicates that water ice could be deficient even outside the sublimation radius.}
%This reduction of water ice may be more significant in the disk around Herbig Ae stars than in the disk around T Tauri star, since the former would have higher dust temperature at a given radius.}
{\item At $r \gtrsim 30$ AU, the cycle decreases the D/H ratios in water ice from $\sim \!\! 2\times10^{-2}$, 
which set by the collapsing core model, to 10$^{-3}$--10$^{-2}$ within 10$^6$ yr, without significant decrease of the column densities. 
The resultant ratio depends on the strength of mixing and the radial distance from the central star.
If $E_{\rm des}$ is 600 K for atomic hydrogen, the D/H ratio of water ice decreases to 10$^{-4}$--10$^{-3}$ at $r \sim 20$--30 AU without 
significantly decreasing the water ice column density.
Our model suggests that the D/H ratio of cometary water could be established (i.e. cometary water could be formed) in the solar nebula, even if the D/H ratio of water ice formed in
the parent molecular cloud/core was very high ($\sim$10$^{-2}$) as observed in some hot corinos.}
{\item We confirmed that water vapor has moderately high abundance ($\gtrsim$10$^{-8}$) in the three regions: (i) the midplane inside the snowline, 
(ii) the inner disk surface, and (iii) the outer disk as shown in previous works. We found that this structure conserves in the model with vertical mixing.
Inside the snow line, the D/H ratios in water vapor become as low as 10$^{-5}$ in the model with mixing.}
{\item The D/H ratios of water vapor and ice are different, especially in the inner regions ($r \lesssim 30$ AU), regardless of the strength of vertical mixing.
It suggests that it is difficult to constrain the D/H ratios of water ice in the inner regions from the observations of D/H ratios of water vapor toward protoplanetary disks.}
\end{enumerate}

\acknowledgments
We are grateful to Prof. Nigel Mason and Dr. Bhala Sivaraman, and Dr. Roland Gredel for providing useful data on photolysis.
H.N. acknowledges SR16000 at YITP in Kyoto University for the numerical calculations.
This work was partly supported by Grant-in-Aids for Scientific Research, 21244021, 21740137, 23103004, 23103005, 23540266, and 25400229, 
and Global COE programs hFoundation of International Center for Planetary Sciencehand gThe Next Generation of Physics, 
Spun from Universality and Emergenceh of the Ministry of Education, 
Culture, Sports, Science and Technology of Japan (MEXT).
F.H. and V.W. are grateful to the French CNRS/INSU PCMI for its financial support.
K.F. is supported by the Research Fellowship from the Japan Society for the Promotion of Science (JSPS) for Young Scientists.
Some kinetic data we used have been downloaded from the online database KIDA (http://kida.obs.u-bordeaux1.fr).

\clearpage

%% the file, uncomment the \plotone, \plottwo, and
%% If you need a layout that cannot be achieved with \plotone or
%% \plottwo, you can invoke the graphicx package directly with the
%% \includegraphics command or use \plotfiddle. For more information,
%% please see the tutorial on "Using Electronic Art with AASTeX" in the
%% documentation section at the AASTeX Web site,
%% http://www.journals.uchicago.edu/AAS/AASTeX.
%%
%% The examples below also include sample markup for submission of
%% supplemental electronic materials. As always, be sure to check
%% the instructions to authors for the journal you are submitting to
%% for specific submissions guidelines as they vary from

%% This example uses \plotone to include an EPS file scaled to
%% 80% of its natural size with \epsscale. Its caption
%% has been written to indicate that additional figure parts will be
%% available in the electronic journal.

%\begin{figure}
%\plotone{f4_new.eps}
%\caption{D/H ratios of selected species in the model with $\alpha_{z} = 0$ (thin lines), and with $\alpha_{z} = 10^{-2}$ (thick lines) as functions of vertical column density at $r = 10$ AU (left panels) and 30 AU (right panels) at $t = 1$ Myr. The solid lines represent gas-phase species, while the dashed lines represent ice-mantle species.\label{fig:vd_dh}}
%\end{figure}

%\begin{figure}
%\epsscale{0.7}
%\plotone{f5.eps}
%\caption{Temporal variation of abundances of selected species in the model with $\alpha_z = 10^{-2}$ at the midplane of $r=10$ AU (upper panel) and 30 AU (lower panel). The solid lines represent gas-phase species, while the dashed lines represent ice-mantle species. \label{fig:tevol}}
%\end{figure}

%\begin{figure}
%\epsscale{0.6}
%\rotatebox{90}{\plotone{fig10.eps}}
%\caption{Spatial distributions of HDO$_{{\rm ice}}$/H$_2$O$_{{\rm ice}}$ ratio in the model with $\alpha_z$ = 0 (left panels), 10$^{-3}$ (middle panels), and 10$^{-2}$ (right panels) at 10$^5$ yr (top panels) and 10$^{6}$ yr (bottom panels). The vertical axes represent height normalized by the radius. The dashed lines indicate vertical visual extinction of unity and ten. The values are only shown in the regions where the water ice abundance is higher than 10$^{-12}$. \label{fig:2d_h2o_ice_dh}}
%\end{figure}

%\clearpage

%\begin{figure}
%\epsscale{0.8}
%\plotone{f14.eps}
%\caption{Temporal variations of the ratio of the product rate of HDO ice to that of H$_2$O ice at 10$^4$ yr (red), 10$^5$ yr (green), and 10$^6$ yr (red).\label{fig:phdo/ph2o}}
%\end{figure}

\clearpage

%\begin{figure}
%\epsscale{0.6}
%\plotone{fig19.eps}
%\caption{Radial variations of water ice column density at $t=10^5$ yr in the model with (dashed) and without (dotted) dead zone. The solid line indicates the values 
%in the model without mixing. \label{fig:dz_r}}
%\end{figure}

%\begin{figure}
%\epsscale{0.6}
%\plotone{f12.eps}
%\caption{The HDO/H$_2$O ratio relative to the HD/H$_2$ ratio as functions of gas temperature. \label{fig:fvst}}
%\end{figure}

\renewcommand\thefootnote{\alph{footnote}}
\begin{table}
\begin{center}
\caption{Observed HDO/H$_2$O Ratios Toward Class 0--I Sources.\label{table:hdo}}
\begin{tabular}{ccc}
\hline\hline
Source             & \multicolumn{2}{c}{HDO/H$_2$O}                                        \\ 
                   & Single Dish                            & Interferometer               \\
\hline
IRAS 16293-2422    & (1.4--5.8)$\times10^{-2}$\footnotemark[1] & (0.66--1.2)$\times10^{-3}$\footnotemark[2]    \\
NGC 1333-IRAS2A    & $\geq$10$^{-2}$\footnotemark[3]           & (0.3--8)$\times 10^{-2}$\footnotemark[4]        \\
NGC 1333-IRAS4A    & --                                        & (0.5--3)$\times 10^{-2}$\footnotemark[4]              \\
NGC 1333-IRAS4B    & --                                        & $\lesssim$6$\times10^{-4}$\footnotemark[5]   \\
%L1157 B-1          & (0.4--2)$\times10^{-3}$\footnotemark[6]   & --                           \\
\hline
\end{tabular}
\tablenotetext{a}{\citet{coutens12}; the value at $T > 100$ K in their modeling.}
\tablenotetext{b}{\citet{persson13}; Source A.}
\tablenotetext{c}{\citet{liu11}; the value at $T > 100$ K in their modeling.}
\tablenotetext{d}{\citet{taquet13a}.}
\tablenotetext{e}{\citet{jorgensen10}.}
%\tablenotetext{f}{\citet{codella12}.}
\end{center}
\end{table}

\renewcommand\thefootnote{\arabic{footnote}}

\begin{table}
\begin{center}
\caption{Formation Reactions of Water and Its Isotopologues at High Temperatures. \label{table:react}}
%\footnotesize
%\scriptsize
\begin{tabular}{cccccccccccccc}
\hline\hline
        &   &      &   Reaction      &        &   &   &$\alpha$ & $\beta$ & $\gamma$ & Reference\\
\hline
O       & + &H$_2$ &   $\rightarrow$ & OH     & + & H & 3.44(-13) & 2.67 & 3.16(3) & 1\\
O       & + &HD    &   $\rightarrow$ & OH     & + & D & 9.01(-13) & 1.90 & 3.73(3) & 2\\
        &   &      &   $\rightarrow$ & OD     & + & H & 1.57(-12) & 1.70 & 4.64(3) & 2\\
O       & + &D$_2$ &   $\rightarrow$ & OD     & + & D & 1.57(-12) & 1.70 & 4.64(3) & 3\\
OH      & + &H$_2$ &   $\rightarrow$ & H$_2$O & + & H & 2.05(-12) & 1.52 & 1.74(3) & 1\\
OH      & + &HD    &   $\rightarrow$ & H$_2$O & + & D & 2.12(-13) & 2.70 & 1.26(3) & 4\\
        &   &      &   $\rightarrow$ & HDO    & + & H & 6.00(-14) & 1.90 & 1.26(3) & 4\\
OH      & + &D$_2$ &   $\rightarrow$ & HDO    & + & D & 3.24(-13) & 2.73 & 1.58(3) & 5\\
OD      & + &H$_2$ &   $\rightarrow$ & HDO    & + & H & 2.05(-12) & 1.52 & 1.74(3) & 3\\
OD      & + &HD    &   $\rightarrow$ & HDO    & + & D & 2.12(-13) & 2.70 & 1.26(3) & 3\\
        &   &      &   $\rightarrow$ & D$_2$O & + & H & 6.00(-14) & 1.90 & 1.26(3) & 3\\
OD      & + &D$_2$ &   $\rightarrow$ & D$_2$O & + & D & 3.24(-13) & 2.73 & 1.58(3) & 3\\
\hline
\end{tabular}
\tablecomments{The rate coefficient is calculated as $k=\alpha(T_{\rm g}/300)^{\beta}\exp(-\gamma/T_{\rm g})$, where $T_{\rm g}$ is gas temperature.}
\tablenotetext{1}{UMIST database \citep{woodall07}.}
\tablenotetext{2}{\citet{bergin99}.}
\tablenotetext{3}{\citet{talukdar96} found that the reaction rate of OD + H$_2$ $\rightarrow$ HDO + H is the same as that of OH + H$_2$ $\rightarrow$ H$_2$O + H, 
while the rate of OD + D$_2$ $\rightarrow$ D$_2$O + D is the same as that of OH + D$_2$ $\rightarrow$ HDO + D in their experiments. 
These results may be interpreted as meaning that the reaction rates depends on whether the reaction involves H atom or D atom abstraction \citep[see also ][]{oba12}. 
We assume that the reaction rate of O + D$_2$ $\rightarrow$ OD + D is the same as that of O +HD $\rightarrow$ OD + H.
We also assume that the rate of OD + HD $\rightarrow$ HDO + D is the same as that of OH + HD $\rightarrow$ H$_2$O + D, 
while the rate of OD + HD $\rightarrow$ D$_2$O + H is the same as that of OH + HD $\rightarrow$ HDO + H.}
\tablenotetext{4}{\citet{talukdar96}.}
\tablenotetext{5}{Fit to experimental results of \citet{talukdar96} in NIST Chemical Kinetics Database (http://kinetics.nist.gov/kinetics/).}
\end{center}
\end{table}

\begin{table}
\begin{center}
\caption{Desorption Energy of Selected Species.\label{table:edes}}
\begin{tabular}{cccc}
\tableline\tableline
Species & $E_{\rm des}$ [K]  \footnotemark[1] & Species & $E_{\rm des}$ [K] \\
\tableline
H$_2$  & 430   & NO     & 1600 \\
H      & 450   & HNO    & 2050 \\
H$_2$O & 5700  & CO     & 1150 \\
OH     & 2850  & CO$_2$ & 2650 \\
O      & 800   & H$_2$CO& 2050 \\
O$_2$  & 1000  &        &      \\
\tableline
\end{tabular}
\tablenotetext{1}{The values are taken from \citet{garrod06}.}
%\tablecomments{(a) $a(-b)$ means $a\times10^{-b}$.}
%% Any table notes must follow the \end{tabular} command.
\end{center}
\end{table}

\begin{table}
\begin{center}
\caption{Initial Abundances of Selected Species for Our Disk Models.\label{table:initial}}
\begin{tabular}{cccc}
\tableline\tableline
Species & Abundance\footnotemark[1] & Species & Abundance \\
\tableline
H$_2$  & 5.0(-1)   & NH$_3$ & 1.4(-5)\\
HD     & 8.0(-6)   & NH$_2$D& 5.1(-7)\\
D$_2$  & 9.6(-7)   & H$_2$CO& 1.2(-5)\\
H      & 3.9(-5)   & HDCO   & 3.0(-7)\\
D      & 9.0(-7)   & CO     & 3.6(-5)\\
H$_2$O & 1.2(-4)   & CO$_2$ & 3.5(-6)\\
HDO    & 2.3(-6)   & O$_2$  & 8.3(-9)\\
CH$_4$ & 1.5(-5)   & O      & 1.5(-12)\\
CH$_3$D& 5.8(-7)   &        &         \\
\tableline
\end{tabular}
\tablenotetext{1}{$a(-b)$ means $a\times10^{-b}$.}
%\tablecomments{(a) $a(-b)$ means $a\times10^{-b}$.}
%% Any table notes must follow the \end{tabular} command.
\end{center}
\end{table}

\end{document}